\documentclass[twocolumn]{aa}

\usepackage{graphicx}
\usepackage[varg]{txfonts}
\usepackage{xcolor}
\usepackage{verbatim}
\usepackage{textcomp, gensymb}
\usepackage{acronym}
\usepackage{natbib}
\usepackage[colorlinks=true,citecolor=blue]{hyperref} 
\usepackage{tablefootnote}
\usepackage{tabularx,booktabs}
\usepackage{makecell}
\usepackage[normalem]{ulem} 
\usepackage{letltxmacro}
\usepackage{subcaption}

\bibpunct{(}{)}{;}{a}{}{,} 

\newcolumntype{C}{>{\centering\arraybackslash}X}
\newcolumntype{L}{>{\raggedright\arraybackslash}X}
\newcolumntype{R}{>{\raggedleft\arraybackslash}X}

\acrodef{BH}{black hole}
\acrodef{BBH}{binary black hole}
\acrodef{SN}{supernova}
\acrodefplural{SN}[SNe]{supernovae}
\acrodef{PPISN}{pulsational-pair instability supernova}
\acrodefplural{PPISN}[PPISNe]{pulsational-pair instability supernovae}
\acrodef{NS}{neutron star}
\acrodef{GWTC-4}{the fourth Gravitational-Wave Transient Catalog}
\acrodef{GWTC-5}{the fifth Gravitational-Wave Transient Catalog}
\acrodef{CHE}{chemically homogeneous evolution}
\acrodef{LVK}{LIGO-Virgo-KAGRA}
\acrodef{AM}{angular momentum}
\acrodef{MS}{main sequence}
\acrodef{LBV}{luminous blue-variable}
\acrodef{SFH}{star formation history}
\acrodef{CO}{compact object}
\acrodef{DCO}{double compact object}
\acrodef{CE}{common envelope}
\acrodef{CEE}{common envelope evolution}
\acrodef{MT}{mass transfer}
\acrodef{SMT}{stable mass transfer}
\acrodef{CHE}{chemically homogeneous evolution}
\acrodef{PDF}{probability distribution function}
\acrodef{CDF}{cumulative distribution function}
\acrodef{KDE}{kernel density estimate}
\acrodef{CI}{confidence interval}

\newcommand{\Msun}{\ensuremath{\xspace M_{\odot}}\xspace}

\newcommand{\Zsun}{\ensuremath{\xspace Z_{\odot}}\xspace}
\newcommand{\ZsunByTen}{\ensuremath{\xspace Z_{\odot}/10}\xspace}
\newcommand{\ZsunByFifty}{\ensuremath{\xspace Z_{\odot}/50}\xspace}
\newcommand{\kms}{\ensuremath{\xspace \mathrm{km}\;\mathrm{s}^{-1}}\xspace}
\newcommand{\MOne}{\ensuremath{\xspace M_1}\xspace}
\newcommand{\MTwo}{\ensuremath{\xspace M_2}\xspace}
\newcommand{\MThree}{\ensuremath{\xspace M_3}\xspace}
\newcommand{\Mco}{\ensuremath{\xspace M_{\mathrm{CO}}}\xspace}
\newcommand{\Mchirp}{\ensuremath{\xspace \mathcal{M}}\xspace}
\newcommand{\MchirpS}{\ensuremath{\xspace \mathcal{M}_{\mathrm{S}}}\xspace}
\newcommand{\dNdMS}{\ensuremath{\xspace \mathrm{d}N/\mathrm{d}\MchirpS}\xspace}

\newcommand{\Mns}{\ensuremath{\xspace M_{\mathrm{NS}}}\xspace}
\newcommand{\Mbh}{\ensuremath{\xspace M_{\mathrm{BH}}}\xspace}
\newcommand{\MbhOne}{\ensuremath{\xspace M_{\mathrm{BH,1}}}\xspace}
\newcommand{\MbhTwo}{\ensuremath{\xspace M_{\mathrm{BH,2}}}\xspace}
\newcommand{\qbbh}{\ensuremath{\xspace q_{\mathrm{BBH}}}\xspace}
\newcommand{\Pbbh}{\ensuremath{\xspace P_{\mathrm{BBH}}}\xspace}
\newcommand{\Tinsp}{\ensuremath{\xspace T_\mathrm{insp}}\xspace}
\newcommand{\dv}{\ensuremath{\xspace \mathrm{d}}\xspace}
\newcommand{\aSF}{\ensuremath{\xspace a_{\mathrm{SF}}}\xspace}
\newcommand{\bSF}{\ensuremath{\xspace b_{\mathrm{SF}}}\xspace}
\newcommand{\cSF}{\ensuremath{\xspace c_{\mathrm{SF}}}\xspace}
\newcommand{\dSF}{\ensuremath{\xspace d_{\mathrm{SF}}}\xspace}
\newcommand{\mOne}{\ensuremath{\xspace m_{1}}\xspace}
\newcommand{\mTwo}{\ensuremath{\xspace m_{2}}\xspace}

\newcommand{\Mprog}{\ensuremath{\xspace M_{\mathrm{prog}}}\xspace}
\newcommand{\ffb}{\ensuremath{\xspace f_{\mathrm{fb}}}\xspace}
\newcommand{\betaTherm}{\ensuremath{\xspace \beta_{\mathrm{th}}}\xspace}
\newcommand{\tHub}{\ensuremath{\xspace T_{\mathrm{H}}}\xspace}
\newcommand{\fGamma}{\ensuremath{\xspace f_{\gamma}}\xspace}
\newcommand{\fGammaStar}{\ensuremath{\xspace f_{\gamma,*}}\xspace}
\newcommand{\fGammaCO}{\ensuremath{\xspace f_{\gamma,\mathrm{BH}}}\xspace}
\newcommand{\alphaCE}{\ensuremath{\xspace \alpha_{\mathrm{CE}}}\xspace}
\newcommand{\cth}{\ensuremath{\xspace C_{\mathrm{th}}}\xspace}
\newcommand{\pastro}{\ensuremath{\xspace p_{\mathrm{astro}}}\xspace}

\newcommand{\modelMaltsevBalanced}{\textsf{Maltsev 25: Balanced}\xspace}
\newcommand{\modelMaltsevOptimistic}{\textsf{Maltsev 25: Optimistic}\xspace}
\newcommand{\modelMaltsevPessimistic}{\textsf{Maltsev 25: Pessimistic}\xspace}
\newcommand{\modelFryerTwelveRapid}{  \textsf{Fryer 12: Rapid}\xspace}
\newcommand{\modelFryerTwelveDelayed}{\textsf{Fryer 12: Delayed}\xspace}
\newcommand{\modelFryerTwentyTwo}{    \textsf{Fryer 22}\xspace}
\newcommand{\modelMullerMandel}{      \textsf{Mandel \& Muller 20}\xspace}
\newcommand{\modelBrcekCores}{$M_\mathrm{core}$: Brcek25\xspace}
\newcommand{\modelHurleyCores}{$M_\mathrm{core}$: Hurley00\xspace}

\newcommand{\modelTwoStage}{Two Stage\xspace}

\newcommand{\modelAccEffCOne}{$C_\mathrm{th}=1$\xspace}
\newcommand{\modelAccEffCHundred}{$C_\mathrm{th}=100$\xspace}

\defcitealias{Maltsev_etal.2025_ExplodabilityCriteriaNeutrinodriven}{M25} 

\usepackage{tikz,xcolor,hyperref}
\definecolor{lime}{HTML}{A6CE39}
\DeclareRobustCommand{\orcidicon}{\hspace{-2mm}
	\begin{tikzpicture}
	\draw[lime, fill=lime] (0,0)
	circle [radius=0.16]
	node[white] {{\fontfamily{qag}\selectfont \tiny \,ID}};
	\draw[white, fill=white] (-0.0525,0.095)
	circle [radius=0.007];
	\end{tikzpicture}
	\hspace{-3mm}
}
\newcommand{\orcid}[1]{\href{https://orcid.org/#1}{\orcidicon}}

\begin{document} 

\title{
Good things always come in 3s: Trimodality in the binary black-hole chirp-mass distribution supports bimodal black-hole formation
} 

\author{ R.\ Willcox 
  \orcid{0000-0003-1817-3586} 
  \inst{\ref{inst:kul_ivs},\ref{inst:lgi}}
  \thanks{\href{mailto:reinhold.willcox@kuleuven.be}{reinhold.willcox@kuleuven.be}} 
\and  
F.\ R.\ N.\ Schneider
  \orcid{0000-0002-5965-1022}
  \inst{\ref{inst:hits},\ref{inst:uheidel1}}
\and  
E. Laplace
  \orcid{0000-0003-1009-5691}
  \inst{\ref{inst:kul_ivs},\ref{inst:lgi},\ref{inst:uams},\ref{inst:hits}}
\and
P. Podsiadlowski
\orcid{0000-0002-8338-9677}
  \inst{\ref{inst:lcsa},\ref{inst:uoxford},\ref{inst:hits}}
\and
K. Maltsev
  \orcid{0000-0001-8060-7416}
  \inst{\ref{inst:mpia}}
\and
\mbox{
I. Mandel
  \orcid{0000-0002-6134-8946}
  \inst{\ref{inst:monash},\ref{inst:ozgrav}}
  }
\and
P.\ Marchant
  \orcid{0000-0002-0338-8181} 
  \inst{\ref{inst:gent}}
\and  
H. Sana
  \inst{\ref{inst:kul_ivs},\ref{inst:lgi}}
\and
T. G. F. Li
  \inst{
  \ref{inst:kul_itp},\ref{inst:kul_dee},\ref{inst:lgi}}
\and
T. Hertog
  \orcid{0000-0002-9021-5966}
  \inst{\ref{inst:kul_itp},\ref{inst:lgi}}
}

\institute{
    {Institute of Astronomy, KU Leuven, Celestijnenlaan 200D, 3001 Leuven, Belgium\label{inst:kul_ivs}} 
    \and 
    {Leuven Gravity Institute, KU Leuven, Celestijnenlaan 200D, box 2415, 3001 Leuven, Belgium \label{inst:lgi}}
    \and
    {Heidelberger Institut f\"ur Theoretische Studien, Schloss-Wolfsbrunnenweg 35, 69118 Heidelberg, Germany \label{inst:hits}} 
    \and
    {Astronomisches Rechen-Institut, Zentrum f\"ur Astronomie der Universit\"at Heidelberg, M\"onchhofstr. 12-14, 69120 Heidelberg, Germany \label{inst:uheidel1}}
    \and
    {Anton Pannekoek Institute of Astronomy, University of Amsterdam, Science Park 904, 1098 XH Amsterdam, The Netherlands \label{inst:uams}}
    \and
    {London Centre for Stellar Astrophysics, Vauxhall, London, UK \label{inst:lcsa}} 
    \and
    {University of Oxford, St Edmund Hall, Oxford, OX1 4AR, UK \label{inst:uoxford}}
    \and
    {Max-Planck-Institut f\"ur Astronomie, K\"onigstuhl 17, 69117 Heidelberg, Germany \label{inst:mpia}}
    \and
    {School of Physics and Astronomy, Monash University, Clayton, Victoria 3800, Australia \label{inst:monash}}
    \and 
    {OzGrav: The ARC Centre of Excellence for Gravitational Wave Discovery, Hawthorn, VIC 3122, Australia \label{inst:ozgrav}}
    \and
    {Sterrenkundig Observatorium, Universiteit Gent, Krijgslaan 281 S9, 9000 Gent, Belgium \label{inst:gent}}
    \and
    {Institute for Theoretical Physics, KU Leuven, Celestijnenlaan 200D, 3001 Leuven, Belgium \label{inst:kul_itp}}
    \and
    {Department of Electrical Engineering (ESAT), STADIUS, KU Leuven, B-3001 Leuven, Belgium \label{inst:kul_dee}}
}
 
\date{Received: Oct 8, 2025; Accepted Aug 24, 2026}

\abstract
   {
    The latest GWTC-5 release from the LIGO–Virgo–KAGRA (LVK) collaboration builds upon the known population of double compact object mergers and reveals new structures in the chirp-mass distribution of merging binary black holes (BBHs). 
    Below 30\,\Msun, the chirp-mass structure is roughly trimodal, with a sharp drop at $\sim10\,\Msun$ and a more tentative dip at $\sim17\,\Msun$; we argue that these features are well explained by recent, detailed stellar evolution models, which predict a bimodal distribution in the birth masses of black holes (BHs).
    This predicted distribution depends only weakly on metallicity and naturally extends to a trimodality in chirp mass, implying universal structural features that can be tested with LVK observations.
    Using a new compact-remnant mass prescription derived from these models, we performed rapid population synthesis simulations to test the robustness of the predicted chirp-mass structure against uncertainties in binary evolution and cosmic star formation history and compared these simulations with the current observational data.
    The trimodal chirp-mass distribution emerges as a robust outcome of the new remnant-mass model, persisting across adopted variations in binary and cosmic physics. 
    In contrast, traditional BH formation models lacking a bimodal BH mass spectrum fail to reproduce the observed structure, especially around the sharp drop at $\sim10\,\Msun$.
    Intriguingly, the central chirp-mass peak, dominated by unequal-mass BBHs, originates from a previously underappreciated formation pathway in which strong luminous blue variable winds suppress any binary interaction before the first BH forms.
    If isolated binary evolution dominates BBH formation below 30\,\Msun, these universal structure features may provide powerful observational constraints on BH formation, binary evolution, and cosmic star formation, and they may also serve as standard sirens for precision cosmology.
   }

   \titlerunning{Chirp-mass trimodality}
   \authorrunning{R. Willcox et al.}
   \maketitle
   
\section{Introduction}
\label{sec:introduction}

With $\sim\,300$ observations of merging double compact objects detected with the \ac{LVK} gravitational-wave observatory network \citep{TheLigoScientificCollaboration_etal.2026_GWTC50IntroductionVersion}, we can now obtain critical constraints on the final stages of massive stellar binaries, which are mostly inaccessible to electromagnetic observing techniques.
These constraints fundamentally rely on models for the formation of merging double compact objects that attempt to explain both the rates and properties of the observed population.
During the past decade, many such models have been proposed to address this challenge, including dynamical formation, isolated evolution of binary or higher-order multiplicity systems, and binary inspirals in active galactic nuclei, among other more exotic channels \citep[see e.g.,][]{Mapelli.2020_BinaryBlackHole, Zevin_etal.2021_OneChannelRule, Mandel_Farmer.2022_MergingStellarmassBinary, Wong_etal.2021_JointConstraintsFieldcluster}.

Here, we focus on isolated binary evolution by modeling the lifetimes of pairs of gravitationally bound, massive $M \gtrsim 8\,\Msun$ stars from birth, through potentially multiple \ac{MT} episodes, until the violent collapse of the stellar cores, which are sometimes accompanied by \ac{SN} explosions \citealp{Jerkstrand_etal.2026_CorecollapseSupernovae}, and the formation of compact objects, including \acp{BH} and \acp{NS} \citep{Heger_etal.2003_HowMassiveSingle, Heger_etal.2023_BlackHolesEnd}.
We used rapid population synthesis to simulate the evolution of massive binary stars and obtain the predicted chirp mass distribution of merging \acp{BBH}, where the chirp mass,
\begin{equation}
    \Mchirp  = \frac{(\MbhOne\MbhTwo)^{3/5}}{(\MbhOne+\MbhTwo)^{1/5}},
\end{equation}
is a well measured combination of the two \ac{BH} component masses, \MbhOne and \MbhTwo.
Rapid population synthesis allowed us to explore the impact of a wide variety of stellar, binary, and cosmic physics variations. 
We are particularly interested in the treatment of the core-collapse event and the resulting \ac{BH} masses.

Core-collapse events are intrinsically complex hydrodynamical processes that can take months to simulate, making them impractical for exploring the progenitor parameter space at scale \citep{Janka_etal.2007_TheoryCorecollapseSupernovae, Burrows_Vartanyan.2021_CorecollapseSupernovaExplosion, Janka.2025_LongtermMultidimensionalModels}.
Efforts to distill core-collapse outcomes into simplified explodability criteria for use in population modeling have been an ongoing challenge for decades
\citep[e.g., ][]{Fryer_Kalogera.2001_TheoreticalBlackHole, Belczynski_etal.2008_CompactObjectModeling,Fryer_etal.2012_CompactRemnantMass,Pejcha_Thompson.2015_LandscapeNeutrinoMechanism, Muller_etal.2016_SimpleApproachSupernova}.
These criteria often revolve around the final CO-core mass \Mco (or, in some cases, the final He-core mass) because it is a more reliable predictor of core-collapse outcomes than the final total mass, which may be significantly affected by stellar winds or binary mass loss.
Many of these criteria invoke the compactness parameter $\xi_M$ at the onset of core collapse, 
defined as
\begin{equation}
    \xi_M = \frac{M/\Msun}{R(M)/1000\,\mathrm{km}},
\end{equation}
where $M$ is a chosen mass coordinate within a star, and $R(M)$ is the radius at that mass coordinate \citep{OConnor_Ott.2011_BlackHoleFormation}.

Cores with a high compactness are more tightly bound, which may indicate that they are more difficult to explode and thus more likely to leave behind a \ac{BH} 
\citep{OConnor_Ott.2011_BlackHoleFormation, Ugliano_etal.2012_ProgenitorexplosionConnectionRemnant, Ertl_etal.2016_TwoparameterCriterionClassifying, Sukhbold_etal.2016_CorecollapseSupernovae9, Heger_etal.2023_BlackHolesEnd}.
Meanwhile, low-compactness core-collapse progenitors are less tightly bound and thus more likely to lead to successful \ac{SN} explosions, often resulting in \acp{NS}. 
If the explosion energy is insufficient to unbind the entire envelope, some of these layers may fall back onto the newborn \ac{NS}, resulting in a fallback \ac{BH} \citep{Janka.2012_ExplosionMechanismsCorecollapse}.

Previous studies support the idea that compactness and other characteristic stellar structure variables at the onset of core collapse follow a pattern consisting of two or more peaks as a function of the initial or core mass
\citep[e.g., ][]{Sukhbold_etal.2016_CorecollapseSupernovae9,Sukhbold_Woosley.2014_CompactnessPresupernovaStellar,Limongi_Chieffi.2018_PresupernovaEvolutionExplosive, Patton_Sukhbold.2020_RealisticExplosionLandscape,Schneider_etal.2021_PresupernovaEvolutionCompactobject,Takahashi_etal.2023_MonotonicityCoresMassive}.
Recently, a series of papers have argued that features in the late-stage evolution of massive stars lead to a bimodal $\xi_M-\Mco$ relation, with a narrow peak around 7\,\Msun before rising to a broader plateau around 14\,\Msun \citep{Schneider_etal.2021_PresupernovaEvolutionCompactobject, Schneider_etal.2023_BimodalBlackHole, Temaj_etal.2024_ConvectivecoreOvershootingFinal, Laplace_etal.2025_ItsWrittenMassive}. 
The physical origin of these peaks is extensively discussed in \citet{Laplace_etal.2025_ItsWrittenMassive}.

\citet{Schneider_etal.2023_BimodalBlackHole} argue that these compactness peaks should result in a bimodality in the distribution of \ac{BH} masses that formed via direct collapse and, consequently, a trimodality in intrinsic \ac{BBH} chirp masses with a relatively weak metallicity dependence.
Using detailed MESA models of stripped stars at \Zsun and \Zsun/10, \citet{Schneider_etal.2023_BimodalBlackHole} demonstrate that this trimodality naturally produces a dearth in the chirp-mass distribution between $\sim10-12\,\Msun$, particularly if the middle of the three peaks is suppressed evolutionarily, which is consistent with the gap in the observed chirp-mass landscape from the third gravitational-wave transient catalog \citep{Abbott_etal.2023_PopulationMergingCompact}.

The explodability also differs for single stars and binary-stripped stars; for the same core mass, the removal of envelope material makes stripped stars easier to explode  
\citep{Timmes_etal.1996_NeutronStarBlack, Wellstein_Langer.1999_ImplicationsMassiveClose, Brown_etal.2001_FormationHighMass, Podsiadlowski_etal.2004_BinariesCompactComponents, Woosley.2019_EvolutionMassiveHelium,  Laplace_etal.2021_DifferentCorePresupernovaa,  Schneider_etal.2021_PresupernovaEvolutionCompactobject, Schneider_etal.2023_BimodalBlackHole}.
In spite of this difference in explodability outcomes, the core-collapse models used in rapid population synthesis codes have thus far not distinguished between single- and binary-stripped core-collapse progenitors, instead treating them using the same core-mass-based final-fate formalism.

Building on these results, \citealt{Maltsev_etal.2025_ExplodabilityCriteriaNeutrinodriven} (henceforth \citetalias{Maltsev_etal.2025_ExplodabilityCriteriaNeutrinodriven}) introduced a new set of explodability criteria for the neutrino-driven \ac{SN} mechanism, which are calibrated to predictions of a semi-analytical \ac{SN} model \citep{Muller_etal.2016_SimpleApproachSupernova}.
The criteria consider several stellar structure variables at the onset of Fe-core infall to anticipate the final fate of the star.
The authors then used these criteria and the single- and binary-stripped star models from \citet{Schneider_etal.2021_PresupernovaEvolutionCompactobject, Schneider_etal.2023_BimodalBlackHole} to construct a core-collapse \ac{SN} recipe applicable for rapid binary population synthesis, based on the CO-core mass $M_\mathrm{CO}$, metallicity $Z$ and mass loss history.
The recipe is more optimistic about successful explosions than previous core-collapse models (e.g., \citealt{Fryer_etal.2012_CompactRemnantMass,Mandel_Muller.2020_SimpleRecipesCompact}) and predicts a bimodal \ac{BH} formation landscape as a function of \Mco, which is qualitatively preserved regardless of $Z$ and the \ac{MT} history of the \ac{SN} progenitors. 

Valid concerns from the detailed \ac{SN} modeling community have been raised about the overuse of simplified recipes, particularly those heavily dependent on the core compactness, since the compactness parameter alone is not sufficient to determine the explodability of a star \citep[see, e.g.,][]{Couch_etal.2020_SimulatingTurbulenceaidedNeutrinodriven,Boccioli_etal.2023_ExplosionMechanismCorecollapse,Burrows_etal.2025_ChannelsStellarmassBlack}.
However, it is not possible to capture the full \ac{SN} physics in rapid population synthesis, thus simplified prescriptions are required. 
The \citetalias{Maltsev_etal.2025_ExplodabilityCriteriaNeutrinodriven} explodability model does not use such a simplified treatment but employs the \citet{Muller_etal.2016_SimpleApproachSupernova} explosion model, which takes into account the full stellar structure at core collapse, partially addressing the concerns about the over-dependence on compactness.

In this study, we implemented a bimodal \ac{BH} mass prescription based on the explodability model from \citetalias{Maltsev_etal.2025_ExplodabilityCriteriaNeutrinodriven} into the rapid population synthesis code COMPAS \citep{TeamCOMPAS:Riley_etal.2022_COMPASRapidBinary, TeamCOMPAS:Mandel_etal.2025_RapidStellarBinary} to investigate its impact on the observed chirp-mass distribution of merging \acp{BBH} from the LVK collaboration, including the recent update of \ac{GWTC-5} \citep{TheLigoScientificCollaboration_etal.2026_GWTC50IntroductionVersion}.
Previously, \citet{Disberg_Nelemans.2023_FailedSupernovaeNatural} used a similar approach, with an analytical proxy for population synthesis simulations, to argue for the bimodal \ac{BH} mass model and trimodality in the chirp-mass distribution. 
More recently, \citet{Galaudage_Lamberts.2025_CompactnessPeaksAstrophysical} and \citet{Galaudage.2026_CompactnessPeaksSubpopulations} also argued in favor of these features using a Bayesian inference approach with the gravitational wave data. 
The approach here is the first to use full population synthesis to test the bimodal \ac{BH} mass hypothesis against a wide range of astrophysical-modeling uncertainties; we comment more on the differences and consistencies in these approaches in Sect.~\ref{sec:comparison_to_previous_studies}.

The treatment and presentation of the observational data are provided in Sect.~\ref{sec:observations}.
In Sect.~\ref{sec:methodology}, we describe our implementation of the new prescription, as well as the alternative supernova treatments that we considered.
The intrinsic properties of the \ac{BBH} population that come out of the population synthesis are presented in Sect.~\ref{sec:intrinsic_properties_bbhs}.
In Sect.~\ref{sec:detector_frame_chirp_masses}, we compare our predictions for the observable chirp-mass distribution to the observations from the LVK.
We discuss the implications of these results in Sect.~\ref{sec:discussion} and conclude in Sect.~\ref{sec:conclusion}.

\section{The observed chirp-mass distribution}
\label{sec:observations}

\begin{figure}[t!]
\centering
\includegraphics[width=\columnwidth]{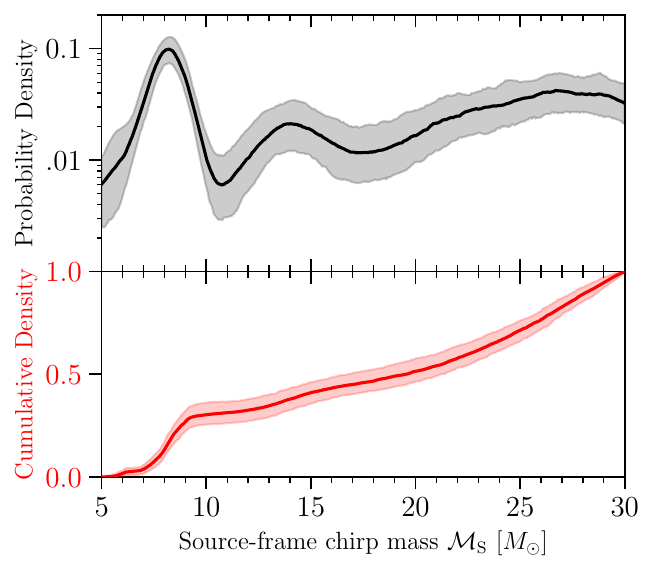}
\caption{
LVK binary black hole chirp-mass distribution, between 5 and 30\,$\Msun$. The probability density distribution (top) shows a strong peak at 8\,\Msun followed by a sharp drop-off at 10\,\Msun, with moderate evidence for additional peaks at 14\,\Msun and 27\,\Msun and a dip at 17\,\Msun, however the appearance of these features is sensitive to the bandwidth of the underlying KDE (see text for details).
The cumulative density distribution (bottom) is unaffected by any choice of bandwidth, so it provides a more reliable representation of the data. 
Here, the transition between the peak and drop above appears as a clear kink at 9\,\Msun, indicating that this particular feature is robust in the data.
}
\label{fig:observed_pdfs}
\end{figure}

The recent \ac{GWTC-5} data release increases the number of confidently observed double compact object mergers, meaning those with false alarm rate < 1\,yr${}^{-1}$ in at least one detection pipeline,
up to 267 \citep{TheLigoScientificCollaboration_etal.2026_GWTC50PopulationProperties}.
In Fig.~\ref{fig:observed_pdfs}, we show the observed source-frame chirp-mass density distribution between 5 and 30\,\Msun, including $\approx180$ \acp{BBH}.
The source-frame chirp mass was calculated as $\MchirpS = \Mchirp_\mathrm{D} / (1+z)$, where the detector-frame chirp mass $\Mchirp_\mathrm{D}$ and redshift $z$ are both relatively well measured in our low-mass region of interest at current detector sensitivity. 
We opted to perform our analysis on the chirp mass instead of the more commonly used component masses because the chirp mass is a directly measured quantity and less subject to modeling uncertainties (see Sect.~\ref{sec:comparison_to_previous_studies} for further discussion). 

We constructed the density distributions using the publicly available source-frame chirp-mass posterior samples from the latest versions of the available datasets,\footnote{GWTC-2.1: \url{https://doi.org/10.5281/zenodo.6513631}}\footnote{GWTC-3: \url{https://doi.org/10.5281/zenodo.8177023}}\footnote{GWTC-4.1: \url{https://doi.org/10.5281/zenodo.20275769}}\footnote{GWTC-5.0: \url{ https://doi.org/10.5281/zenodo.20292639}}
totaling 263 \ac{BBH} merger events with a false alarm rate less than $1\,\mathrm{yr}^{-1}$ and additionally $\pastro>0.5$ (excluding \textsf{GW230630\_070659}, which is thought to be of instrumental origin, \citealt{Soni_etal.2025_LIGODetectorCharacterization}).
For GWTC-2.1 and GWTC-3, we used the cosmologically reweighted datasets to maintain consistency with GWTC-4.1 and GWTC-5.0.
With the removal of the "Mixed" waveform model from the events in GWTC-5.0, there is no longer a waveform model that is common to all reported events. 
We have therefore opted to use a mix of the following waveforms, choosing the first in this list that is available for a given event and which collectively provides a waveform for all our events of interest: \textsf{C00:Mixed}, \textsf{C01:Mixed}, \textsf{C01:IMRPhenomXPHM}, and \textsf{C00:IMRPhenomXPHM-SpinTaylor}.
Different waveforms will result in different posterior samples for a given event, including for the inferred chirp masses.
We do not expect these differences to substantially modify our conclusions, however a quantitative assessment of the impact of different waveform models is beyond the scope of this study.

We subsequently constructed mock populations of merger events by bootstrapping from the posterior samples, drawing with replacement first from the events themselves and subsequently from their posteriors, to avoid over-representing outlier events and to ensure that all events are given equal weighting (since not all provided datasets have equal size).
Each mock population consists of 263 nominal chirp-masses, which are then used to construct an adaptive \ac{KDE} with an initial bandwidth of 0.05\,\Msun.

Compared to a standard \ac{KDE}, an adaptive \ac{KDE} uses a narrower (broader) kernel for more clustered (isolated) samples and does a better job resolving sharp features that result from rapid changes in sample density \citep{Silverman.1986_DensityEstimationStatistics, Wang_Wang.2007_BandwidthSelectionWeighted}. 
The construction of an adaptive \ac{KDE} and the choice of bandwidth are discussed in greater detail in Appendix~\ref{sec:choosing_the_adaptive_kde_bandwidth}.
The median of the adaptive \acp{KDE} from all mock populations is shown in the top plot of Fig.~\ref{fig:observed_pdfs}, together with the 90\% percentile of the range of these \acp{KDE}. The bottom plot similarly shows the median and 90\% percentile of the range of the cumulative density distributions of the mock populations, which are independent of the subjective choice of \ac{KDE} bandwidth.

With the new data, the observed chirp-mass distribution shows a clear peak around 8\,\Msun and a prominent dip at 10\,\Msun, with a corresponding kink at 9\,\Msun in the cumulative density distribution.
While a wider \ac{KDE} bandwidth would wash out this feature in the probability density distribution, a flat, featureless distribution would appear as a straight line in the cumulative density distribution, which is inconsistent with the observed kink, thus the presence of this dip is a robust feature of the data. 
In Appendix~\ref{sec:choosing_the_adaptive_kde_bandwidth}, we argue for a bandwidth no larger than $\approx0.05\,\Msun$ in order to resolve this feature consistently with the cumulative density distribution.

Further, the data show a rise again up to $\sim27\,\Msun$, and there may also be a smaller peak around $14\,\Msun$ with a dearth between $\sim16-20\,\Msun$, although the bootstrapping and the straightness of the cumulative density distribution in this regime indicate that these could be consistent with fluctuations due to small-number statistics.
The observed trimodal structure was first discussed in \citet{Tiwari_Fairhurst.2021_EmergenceStructureBinary}.

\section{Methodology}
\label{sec:methodology}

\subsection{Rapid population synthesis}
\label{sec:rapid_population_synthesis}

To test the predictions of a bimodal \ac{BH} mass distribution on the LVK-observable chirp masses, we used the rapid population synthesis code COMPAS
\citep{Stevenson_etal.2017_FormationFirstThree, Vigna-Gomez_etal.2018_FormationHistoryGalactic,TeamCOMPAS:Riley_etal.2022_COMPASRapidBinary,TeamCOMPAS:Mandel_etal.2025_RapidStellarBinary}.
Analytical fits and parameterizations to detailed stellar and binary evolution physics allow COMPAS to simulate the life of a single binary in a fraction of a second, enabling investigations into the population-level impacts of many uncertain physical parameters, albeit with well known caveats common to many population synthesis codes due to the over-simplified treatment of various physical processes
\citep{deMink_Belczynski.2015_MergerRatesDouble, Giacobbo_Mapelli.2018_ProgenitorsCompactobjectBinaries, Tang_etal.2020_DependenceGravitationalWave, Bavera_etal.2021_ImpactMasstransferPhysics,
Broekgaarden_etal.2021_ImpactMassiveBinary,  Broekgaarden_etal.2022_ImpactMassiveBinary, Mandel_Broekgaarden.2022_RatesCompactObject, Belczynski_etal.2022_UncertainFutureMassive}.

For each model variation explored in this study, we simulated $10^6$ binaries with initial conditions sampled from standard, non-correlated distributions. 
The initial mass of the primary \mOne was drawn from the Kroupa initial mass function,  $\mOne \sim \mOne^{-\alpha_\mathrm{IMF}}$, with $\alpha_\mathrm{IMF}=2.3$, in the range [5,150]\,\Msun \citep{Kroupa.2001_VariationInitialMass}.
The mass of the secondary \mTwo was drawn from a uniform mass ratio distribution, $q = \mTwo/\mOne \sim [0.01,1]$ \citep{Sana_etal.2012_BinaryInteractionDominates,Shenar_etal.2022_TarantulaMassiveBinary}, and the initial separation $a$ from a log-uniform distribution, $\log_{10}(a/\mathrm{AU}) \sim U(-2,3)$ \citep{Opik.1924_StatisticalStudiesDouble}, which implicitly accounts for an interacting binary fraction of $\sim70\%$ \citep{Sana_etal.2012_BinaryInteractionDominates}.
It is possible that the initial parameter distributions are in fact correlated \citep{Moe_DiStefano.2017_MindYourPs}, which may have an impact on the final \ac{BBH} populations. We leave such exploration to a future study.

The metallicity $Z$ of the binary was sampled from a log-uniform distribution, $\log_{10}(Z) \sim U(-4, \log_{10}(0.03))$, however, samples were subsequently reweighted according to the adopted cosmic \ac{SFH}. 
The initial distributions are assumed to be independent of metallicity, which is consistent with recent observations down to SMC metallicity, $Z_\mathrm{SMC} \approx \Zsun/5$ \citep{Sana_etal.2025_HighFractionClose}, where $\Zsun=0.0142$ \citep{Asplund_etal.2009_ChemicalCompositionSun}.

The evolution of both binary components started at the zero-age main sequence and proceeded through potentially multiple phases of \ac{SMT} or \ac{CEE}, until both components evolved into compact remnants, including \acp{BH}, \acp{NS}, or white dwarfs, unless the binary merged prior to that point or the simulation runtime exceeded the age of the universe (roughly the Hubble time \mbox{$\tHub \approx $ 14\,Gyr)} \citep{PlanckCollaboration_etal.2020_Planck2018Results}.
We focused in this study on the formation of \acp{BH} and \acp{BBH}, although the variations explored will certainly impact, among other things, X-ray and inert \ac{BH} binaries, and \ac{NS} populations including merging \ac{BH}-\acp{NS}. 

We are primarily interested in variations related to the \ac{BH} mass function; namely, how the \ac{BH} populations differ between the bimodal \ac{BH} formation mechanism that we use and the alternatives that have been adopted in earlier studies on \ac{BBH} synthesis.
To answer this, we used four \ac{SN} prescriptions that are standard in population synthesis (see Sect.~\ref{sec:alternative_sn_models}) in addition to the new bimodal \ac{BH} model based on \citepalias{Maltsev_etal.2025_ExplodabilityCriteriaNeutrinodriven}, which is described in detail below.
To inspect the robustness of these results, we included additional variations in Sect.~\ref{sec:chirp_mass_predictions_for_all_models} that account for uncertainties in stellar and binary evolution (see Appendix~\ref{sec:default_and_variations_to_the_stellar_and_binary_physics}), as well as variations to the cosmic \ac{SFH} (see Appendix~\ref{sec:cosmic_star_formation_history}).
The stellar and binary evolution variations were paired with each of the \ac{SN} models using the default cosmic \ac{SFH} model, while the cosmic \ac{SFH} variations are applied to each \ac{SN} model using the default choices for the stellar and binary physics.

\subsubsection{Bimodal BH mass model}
\label{sec:bimodal_black_hole_mass_model}

\begin{figure*}[hbt!]
\centering
\includegraphics[width=\textwidth]
{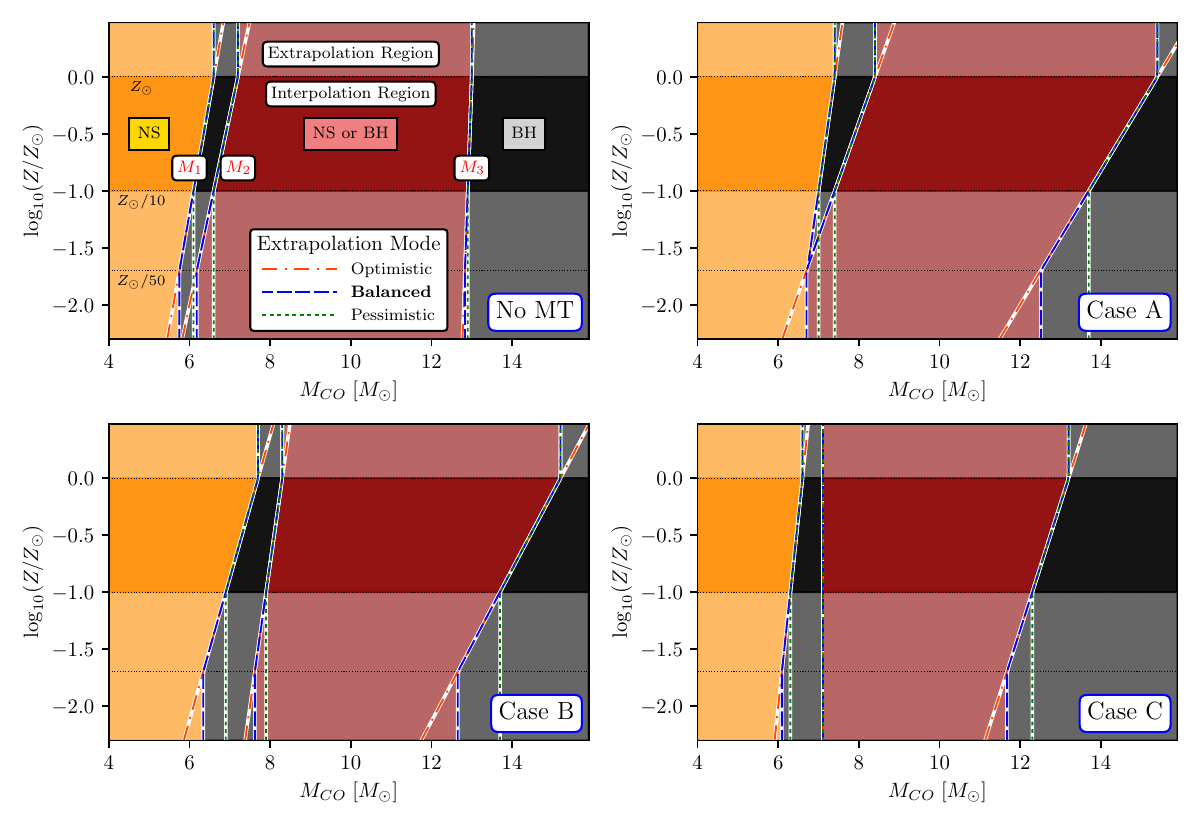}
\caption{
Core-collapse outcome landscape from the bimodal \ac{BH} model. The outcome of core collapse is shown in colors as a function of $Z$, \Mco at He depletion, and \ac{MT} history, for the bimodal \ac{BH} mass models explored in this work. 
MT history refers to the timing (Case A, B, or C) of the first binary interaction in which the core-collapse progenitor was a donor, or `No MT' if the progenitor was never a \ac{MT} donor, shown in the four subplots.
We include three variations to this model, Optimistic, Balanced, and Pessimistic, corresponding to how the model is extrapolated outside of the range $Z \sim [0.1 Z_{\odot}, Z_{\odot}]$.
The interpolation region is highlighted in darker colors to emphasize that this is the region where we are more confident in the models. 
In the extrapolation region, we show in lighter colors the outcomes for only the Balanced extrapolation mode, which we treat as the default.
See text for details.
}
\label{fig:remnant_mass_model_extrapolations}
\end{figure*}

The core-collapse \ac{SN} recipe introduced in \citetalias{Maltsev_etal.2025_ExplodabilityCriteriaNeutrinodriven} was specifically designed only to predict the final fate (failed vs. successful \ac{SN}) and compact remnant type (direct-collapse \ac{BH}, fallback \ac{BH} or \ac{NS}), but not the remnant's mass. 
Below, we summarize the model, couple it to a prescription for \ac{BH} masses, and extrapolate it in metallicity outside the range \mbox{($Z_\odot/10 \leq Z \leq Z_\odot$)} over which it was originally defined.

Our implementation of the resulting \ac{BH} mass models into the COMPAS code is shown in Fig. \ref{fig:remnant_mass_model_extrapolations}. 
\citetalias{Maltsev_etal.2025_ExplodabilityCriteriaNeutrinodriven} defines three threshold mass boundaries $M_1$, $M_2$, and $M_3$ that are used to predict the final fate of the \ac{SN} progenitor, for a given \Mco. 
These boundary values depend on both $Z$ and \ac{MT} history of the star.
Specifically, the \ac{MT} history classification distinguishes whether the collapsing star first interacted with its companion during the \ac{MS} (Case A), after the \ac{MS} but before core He-depletion (Case B), after core He-depletion (Case C), or if the progenitor evolved as an isolated star retaining its hydrogen-rich envelope up to core collapse (No MT). 
Subsequent \ac{MT} interactions were not considered, since the timing of the first interaction is the most pivotal factor in whether later shell burning will proceed, and thus in deciding the position of the compactness peaks.

The determination of \ac{BH} masses is described below and summarized in Table \ref{tabl:bh_mass_model}.
\begin{itemize}
    \item If $\Mco < M_1$, the collapse is a successful explosion, leaving behind a \ac{NS} remnant. 
    \item If $\MOne \leq \Mco < \MTwo$, or if $\Mco > \MThree$, the star experiences a direct collapse resulting in a \ac{BH}. In this case, the \ac{BH} mass is equal to the mass up to and including the He-layer, under the assumption that any remaining H-rich envelope is always ejected.\footnote{There is some support theoretically \citep{Yoon_Cantiello.2010_EvolutionMassiveStars,
    Antoni_Quataert.2023_NumericalSimulationsRandom, 
    Bronner_etal.2025_ExplosionsPulsatingRed} and observationally \citep{Schootemeijer_etal.2026_ConstantUpperLuminosity} that single \ac{BH} progenitors always lose their H-envelopes prior to core-collapse, but this is not so important for us as the \ac{BBH} progenitors always experience binary mass transfer.}
    \item If $M_2 < \Mco < M_3$, the compact remnant type is treated stochastically. We assume that collapse results in a successful explosion and a \ac{NS} remnant with 90\% probability. Otherwise, we assume that the explosion involved partial fallback onto the proto-\ac{NS}, resulting in a fallback \ac{BH}. This corresponds to fallback model B in \citetalias{Maltsev_etal.2025_ExplodabilityCriteriaNeutrinodriven}. 
\end{itemize}
The values of \MOne, \MTwo, and \MThree for $Z=\Zsun$ and $Z=0.1 \Zsun$ were taken from \citetalias{Maltsev_etal.2025_ExplodabilityCriteriaNeutrinodriven}; see Appendix~\ref{sec:details_of_bimodal_prescription}.

For predicting fallback \ac{BH} masses, we define the fallback fraction 
\begin{equation}
        \ffb = \frac{\Mbh - \Mns}{\Mprog-\Mns},
\end{equation}
where \Mprog is the progenitor He-core mass at core collapse, \Mbh is the final mass of the \ac{BH} after fallback, and \Mns is the nominal \ac{NS} mass, which we take throughout to be $\Mns = 1.4\,\Msun$. 
By default we assumed $\ffb = 50\%$ (though see Appendix~\ref{sec:default_and_variations_to_the_stellar_and_binary_physics}).

\begin{table}
\begin{center}
\caption{
Compact remnant mass model.
}
\begin{tabular}{l|l}       
\hline\hline 
$\Mco$ range & Compact remnant mass $M_\mathrm{rem}$ \\
\hline
$\Mco < M_1$ & $M_\mathrm{rem} = M_\mathrm{NS} = 1.4 \, M_\odot$ \\
\hline
$M_1 \leq \Mco \leq M_2,$ & $M_\mathrm{rem} = \Mprog$ \\
or $\,\,\Mco \geq M_3$ & \\ \hline
$M_2 < \Mco < M_3$ & $M_\mathrm{rem}= M_\mathrm{ NS}$ at 90\% probability, else: \\
& $M_\mathrm{rem} = M_\mathrm{NS} + f_\mathrm{fb}(\Mprog - M_\mathrm{NS})$ \\
\hline 
\end{tabular}
\label{tabl:bh_mass_model} 
\end{center}  
\end{table}

Following \cite{Hurley_etal.2002_EvolutionBinaryStars, Schneider_etal.2024_PresupernovaEvolutionFinal, Schneider_etal.2025_StellarMergersCommonenvelope}, we treated accretion onto a \ac{MS} star as rejuvenation up to a \ac{MS} star of a higher mass at the same fractional age.
If the star was never a \ac{MT} donor, it follows the `No \ac{MT}' classification, regardless of any prior accretion.
The current set of models does not account for the structural changes that occur during accretion events post-\ac{MS}, though these can influence the progenitor-remnant connection \citep{Schneider_etal.2024_PresupernovaEvolutionFinal}.
In some cases, a very massive, effectively single star experiences sufficiently high stellar winds -- typically, during a \ac{LBV} phase -- that it strips itself of its outer H-envelope; in such cases, the subsequent shell burning is assumed to be quenched analogously to such quenching from binary stripping, and so these stars were treated just as Case-B binary stripped stars.

Previous works have shown that \acp{BBH} formed in low-metallicity environments $Z \ll \Zsun/10$ compose the dominant contribution to the total number of sources observable via gravitational wave astronomy \citep[e.g.,][]{Belczynski_etal.2010_EffectMetallicityDetection, Stevenson_etal.2017_FormationFirstThree}, although as we show later, this is not necessarily the case for sources primarily detected at low redshift.
In order to predict BBH formation using the \citetalias{Maltsev_etal.2025_ExplodabilityCriteriaNeutrinodriven} explodability model, we extrapolated it outside the [$\ZsunByTen,\Zsun$] range where it was calibrated.
The extrapolation was extended to higher metallicity as well, though this did not influence \ac{BBH} formation significantly.
We define three extrapolation schemes for the bimodal BH mass model to account for the uncertainty in the $Z$ dependence outside these thresholds, when covering $Z\sim[.0001, .03]$ (the full range of the COMPAS simulations).
This implementation is shown in Fig.~\ref{fig:remnant_mass_model_extrapolations}. 

In all cases, we interpolated linearly in $\log_{10}(Z)$ between the boundaries at \Zsun/10 and \Zsun. 
In the Optimistic variant, we extrapolated linearly in $\log_{10}(Z)$, continuing the trend between the boundaries. 
The name Optimistic suggests enhanced formation of higher-mass ($\sim12\,\Msun$) \acp{BH} at low $Z$.
In the Pessimistic variant, we applied a nearest neighbor extrapolation at both boundaries, suppressing the formation of higher-mass, low-$Z$ \acp{BH}.
In the Balanced variant, we extrapolated linearly in $\log_{10}(Z)$ down to $\log_{10}(\Zsun/50)$ (roughly inspired by \mbox{I Zwicky 18}, one of the lowest-metallicity observed galaxies, see \citealp{Vilchez_Iglesias-Paramo.1998_EnhancedHaEmission}), at which point we switched to a nearest-neighbor extrapolation. 
This choice of lower metallicity was ad hoc but allowed us to explore a moderate choice between the two more extreme alternatives while awaiting new stellar models to better understand pre-\ac{SN} core structures at low $Z$. 
Details of the implementation of the bimodal \ac{BH}-mass prescription can be found in Appendix~\ref{sec:details_of_bimodal_prescription}.

Notably, we found that when considering current detector sensitivity and biases, the differences between the extrapolation schemes are negligible:
the contribution from progenitors that would experience different outcomes under the three extrapolation schemes to the total detectable \ac{BBH} number density constitute less than 0.5\% of the total number density.
This may be a consequence of incorrect assumptions on the binding energies of common envelope progenitors \citep{Marchant_etal.2021_RoleMassTransfer,Klencki_etal.2021_ItHasBe}. 
Thus, throughout this article, we only apply the stellar, binary, and cosmic physics variations to the \modelMaltsevBalanced model, and include the \modelMaltsevOptimistic and \modelMaltsevPessimistic models as one-off variations in Sect.~\ref{sec:chirp_mass_predictions_for_all_models} (see also Appendix~\ref{sec:default_and_variations_to_the_stellar_and_binary_physics}).

\subsubsection{Alternative SN models}
\label{sec:alternative_sn_models}

To compare the bimodal prescription against existing alternatives, we include variations based on several well known \ac{SN} prescriptions, including the \modelFryerTwelveRapid and \modelFryerTwelveDelayed variations from \citet{Fryer_etal.2012_CompactRemnantMass}, \modelFryerTwentyTwo from \citet{Fryer_etal.2022_EffectSupernovaConvection}, and \modelMullerMandel from \citet{Mandel_Muller.2020_SimpleRecipesCompact}.
All of these models predict a continuous \ac{BH} mass spectrum and, except for \modelFryerTwelveRapid, no mass gap between \acp{BH} and \acp{NS} which can introduce features at low chirp mass.

\subsection{Calculating the detection rate}
\label{sec:calculating_detector_frame_chirp_masses}

We first computed the merger rate of detectable \acp{BBH}.
This required convolving the intrinsic yield at a given metallicity $Z$ of merging binary black holes per unit source-frame chirp mass per unit star forming mass with cosmic \ac{SFH} $\mathcal{S}(Z, z)$ per unit time per unit comoving volume, which depends on both metallicity $Z$ and redshift $z$.  
For the \ac{SFH}, we followed the parameterized treatment and default choices described in \citet{vanSon_etal.2023_LocationsFeaturesMass}, and refer to that study for details.  
A brief overview is also included in Appendix~\ref{sec:cosmic_star_formation_history}.

The convolution assigned a rate for each simulated \ac{BBH} to merge at a given redshift, accounting for the binary inspiral time. 
The time between the birth of the stellar binary and the birth of the \ac{BBH} is negligible here because the inspiral time is typically much longer than the lifetimes of massive stars, $O(10^6)\,\mathrm{yrs}$.  
We use the \textsf{Planck18} cosmological model  \citep{PlanckCollaboration_etal.2020_Planck2018Results} as implemented in astropy to convert between redshift and lookback time.

We approximated the sensitivity of the gravitational-wave detector network by requiring detectable BBHs to have a signal-to-noise ratio exceeding a threshold of 8 in a single detector, computed using the method described in \citet{Barrett_etal.2018_AccuracyInferencePhysics}. 
For simplicity, we assumed a single typical detector for the entire observing campaign, using the power spectral density dataset SimNoisePSDaLIGOMidHighSensitivityP1200087 from LALSuite 
\citep{TheLigoScientificCollaboration_etal.2018_LVKAlgorithmLibrary}.  A more careful treatment that incorporates the evolving noise power spectral density of each detector, as well as the overlap times when multiple detectors are operating simultaneously, is left to a future study. 
Averaging over sky locations and orientations provided a detection probability for each combination of \ac{BBH} masses and merger redshift.  

We thus computed the differential rate of detectable \ac{BBH} mergers $\mathrm{d}R/\mathrm{d}\MchirpS$ per \Msun per year, for chirp masses \MchirpS in the region of interest between 5 and 30\,\Msun.
To compare directly to the LVK observations, we additionally computed the number density of detections $\mathrm{d}N/\mathrm{d}\MchirpS$ per \Msun, assuming a 1-year total observing duration, accounting for fluctuations due to Poisson statistics.
Systematic uncertainties in astrophysical and cosmological models likely dominate over these sampling uncertainties.

\section{Intrinsic properties of the BBHs}
\label{sec:intrinsic_properties_bbhs}

\begin{figure*}[ht!]
\centering
\includegraphics[width=\textwidth]{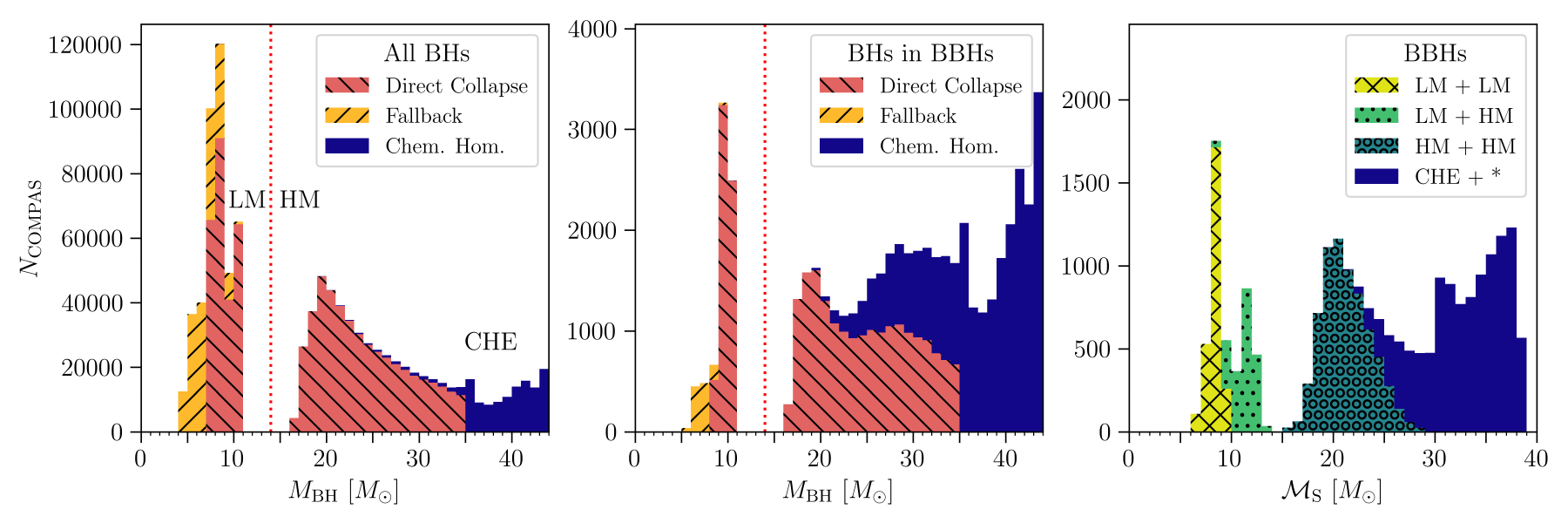}
\caption{
BH and BBH mass distribution from direct collapse, fallback, and chemically homogeneous formation channels, under the \modelMaltsevBalanced model for a simulation of $10^6$ binaries. 
Left panel: distribution of masses of all BHs formed in the simulation. 
Colors and hatching indicate systems which formed via direct collapse (orange) or fallback (yellow), or if they experienced CHE (dark blue), regardless of the explosion mechanism. The vertical dotted line at 14\,\Msun indicates the boundary that we use to distinguish low-mass (LM) and high-mass (HM) BHs, forming from the bimodal BH prescription. 
Middle panel: same as the left plot, but including only components of merging BBHs.
Right panel: BBH source-frame chirp masses \MchirpS.
Colors indicate whether both components are low-mass BHs (LM+LM, lime), a mixture of low- and high-mass BHs (LM+HM, green), both are high-mass BHs (HM+HM, sea green), or if either component formed through CHE (dark blue). 
Bin edges lie at integer values of the ordinate. 
}
\label{fig:mass_spectrum_lmhm}
\end{figure*}

In this section, we focus on the properties of the \ac{BBH} population model from the rapid population synthesis sampled from the log-uniform metallicity distribution, without accounting for any cosmic integration effects or observational biases. 
The distributions show the \acp{BBH} with a merger time less than \tHub following the default choices described at the beginning of Sect.~\ref{sec:rapid_population_synthesis} captured in the default \modelMaltsevBalanced model. 
Thus, in this section, chirp mass refers to this simulation output, prior to convolution with the cosmic \ac{SFH}.

\subsection{The BH and BBH mass distribution}
\label{sec:bh_bbh_mass_spectrum}

In Fig.~\ref{fig:mass_spectrum_lmhm}, we show the masses of \acp{BH} and \acp{BBH} that directly result from the simulated populations under the \modelMaltsevBalanced model. 
The gap in \ac{BH} masses between $\sim11-16\,\Msun$ is immediately apparent. This is a direct result of the assumption of bimodality in the \ac{BH} formation landscape, although the exact locations of the edges of this gap are dependent on the assumed stellar and binary evolution physics (see Sect.~\ref{sec:chirp_mass_predictions_for_all_models}).
We distinguish low-mass (LM) from high-mass (HM) \acp{BH} with a cut at 14\,\Msun, although the precise value is not so important. 

We include an additional category for systems that were flagged as experiencing \ac{CHE} at some point during the simulation, regardless of how they formed. The \ac{CHE} channel occurs when the two binary components are tidally locked at birth and rotate at the orbital period, resulting in efficient mixing and the burning of all available H during the main sequence  
\citep{Marchant_etal.2016_NewRouteMerging, Mandel_deMink.2016_MergingBinaryBlack, deMink_Mandel.2016_ChemicallyHomogeneousEvolutionary}.
\acp{BBH} that formed via \ac{CHE} are more massive, and thus do not affect the features in the lower mass region.

Ignoring \acp{BBH} from the \ac{CHE} channel for the moment, the \ac{BBH} source-frame chirp-mass \MchirpS distribution reveals three clear peaks corresponding to systems formed by two low-mass \acp{BH} (LM+LM), two high-mass \acp{BH} (HM+HM), or one of each (LM+HM).
This trimodality in \MchirpS is an inherent prediction of the bimodal \ac{BH} mass model \citep{Schneider_etal.2023_BimodalBlackHole}.

Serendipitously, the \ac{CHE} binaries produce a sharply rising feature at $\sim30\,\Msun$, which is reminiscent of a peak at this same mass in the LVK observations. 
For the remainder of this article, we omit the \ac{CHE} binaries and focus only on the lower mass regime where their impact is minimized and the effects of the \ac{BH} mass prescription dominate. 

\subsection{Formation of merging BBHs}
\label{sec:formation_of_merging_bbhs}

\begin{figure}[ht!]
\centering
\includegraphics[width=.99\columnwidth]{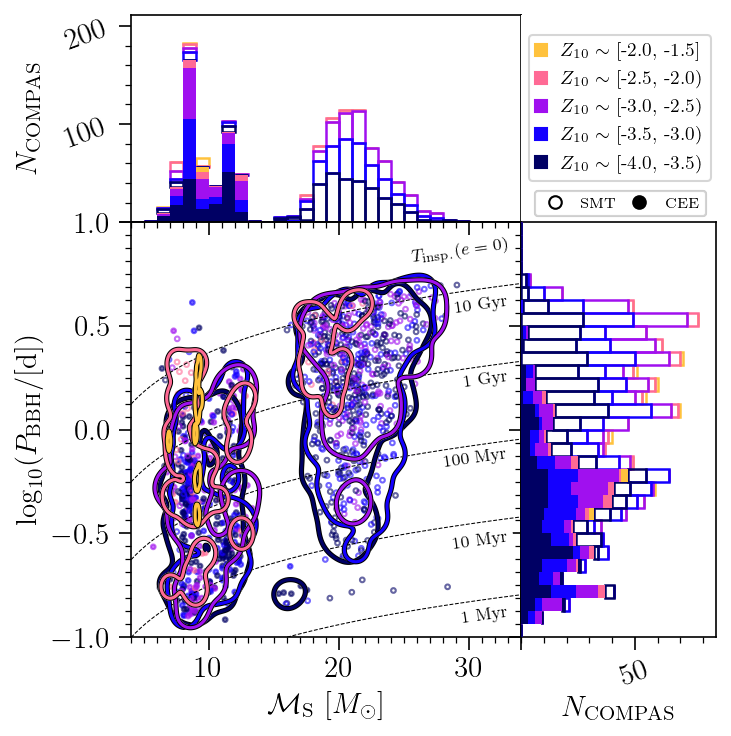}
\caption{
Chirp mass \MchirpS vs. period at \ac{BBH} formation \Pbbh under the default model. 
The chirp mass is the outcome of the initial population synthesis, prior to convolution with a cosmic SFH.
Marginalized histograms of each parameter are shown at the top and right. 
Colors correspond to metallicity bins, with \mbox{$Z_{10}:= \log_{10}(Z)$}. 
Filled scatter points and histograms show binaries that experienced a common envelope event during their evolution; those that experienced only stable mass transfer are depicted in hollow points and histograms (stacked on top of the filled histograms).
Colored contours in the scatter plot bound the regions containing the highest 90\% of the KDE volume for each metallicity bin, determined using a 2D Gaussian adaptive KDE. 
Characteristic inspiral times \Tinsp for circular \acp{BBH} are over-plotted.
Scatter points have been down-sampled to improve clarity, but the histograms show the full output of the simulation. 
}
\label{fig:chirp_mass_vs_P}
\end{figure}

The significant gap around $\MchirpS\sim15\Msun$ observed between the LM+HM and HM+HM peaks (right panel in Fig.~\ref{fig:mass_spectrum_lmhm}) is also present in Fig.~\ref{fig:chirp_mass_vs_P}, where we show the initial period at formation $P_\mathrm{BBH}$ vs. \MchirpS of the merging \acp{BBH}.
Here, we also see a domination of low-metallicity systems among the \ac{BBH} population, as the yields of \acp{BBH} from each metallicity bin become smaller with increasing $Z$.
However, we remind the reader that the initial $Z$-distribution was sampled log-uniformly to underscore the features of the \ac{BBH} that are intrinsic to the population synthesis results, but this distribution is not representative of a realistic cosmic \ac{SFH} and so this over-representation of low-metallicity systems should not be expected to persist after reweighting.

The high mass peak \acp{BBH} (HM+HM) come from the lowest metallicity binaries, covering areas in $P_\mathrm{BBH}$ -- \MchirpS space that contract with increasing metallicity and shift toward longer periods. 
This is an expected consequence of decreasing masses and increasing orbital widening due to increased stellar winds at higher $Z$.
Meanwhile, the \acp{BBH} with the highest metallicity, $\log_{10}Z\sim[-2.0,-1.5]$, only contribute meaningfully to the lowest mass peak (LM+LM).

The two prominent peaks in the marginalized orbital period distribution, around $\Pbbh\approx3\,\mathrm{d}$ and $\approx0.6\,\mathrm{d}$, can be roughly matched to those with inspiral times \Tinsp greater or less than 100\,Myr, respectively.
This divide is primarily driven by the formation channel, with \ac{CEE} binaries dominating at lower periods.
There is also a smaller bump at very short periods, $\Pbbh\sim\mathrm{few}\times0.1\,\mathrm{d}$, the origin of which we leave to a future study.

Intriguingly, the \MchirpS distribution is even more sharply divided by formation channel \citep[as predicted in][]{vandenHeuvel_etal.2017_FormingShortperiodWolfrayet}, with \ac{CEE} binaries populating a significant majority of the lower-mass peaks (LM+LM and LM+HM), but virtually absent from the HM+HM peak. 
This is primarily because the strong \ac{LBV} winds in the HM \ac{BH} progenitors prevent the binaries from reaching the \ac{CEE} phase. 
We discuss the impact of the \ac{LBV} phase in Sect.~\ref{sec:impact_lbv_winds}.
The substantial proportion of \ac{CEE} binaries at the lowest chirp masses is consistent with \citet{vanSon_etal.2022_RedshiftEvolutionBinary} (see their Fig.~4).

\begin{figure}[ht!]
\centering
\includegraphics[width=\columnwidth]{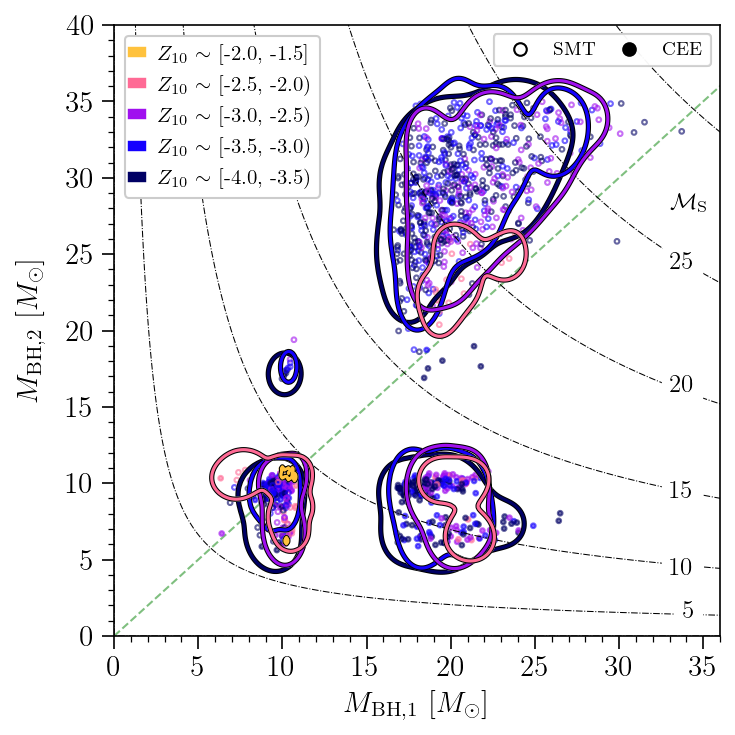}
\caption{
Masses of the primary (\MbhOne) and secondary (\MbhTwo) component in merging \acp{BBH} under the default model.
Primary and secondary are defined at birth, $M_\mathrm{ZAMS,1} > M_\mathrm{ZAMS,2}$, so we directly see which systems experienced mass ratio reversal ($\MbhOne < \MbhTwo$) and which did not. 
Colors correspond to the same metallicity bins as Fig.~\ref{fig:chirp_mass_vs_P}, and filled and hollow points depict systems that went through CEE and SMT formation channels, respectively.
Colored contours bound the regions containing the highest 90\% of the KDE volume for each metallicity bin, determined using a 2D adaptive Gaussian KDE.
The green dashed line shows $\MbhOne=\MbhTwo$.
Black dot-dashed lines show curves of constant chirp mass \MchirpS (in \Msun).
}
\label{fig:m1_vs_m2}
\end{figure}

From the individual component masses (Fig.~\ref{fig:m1_vs_m2}), we found that the low-\MchirpS peak (LM+LM) corresponds to roughly equal mass binaries composed of $\sim10\,\Msun$ components, as expected, while the middle \MchirpS peak (LM+HM) is composed of $\sim10+20\,\Msun$ \acp{BBH}, where the $20\,\Msun$ \ac{BH} came from the initially more massive progenitor star. 
The high \MchirpS peak (HM+HM), by contrast, is composed almost exclusively of binaries that experienced mass ratio reversal, consistent with earlier studies on mass ratio reversal \citep{Broekgaarden_etal.2022_SignaturesMassRatio}.
The mass ratio distribution for each pairing of LM and HM \acp{BH} is shown in Appendix~\ref{sec:bbh_mass_ratios}, wherein each pairing clearly maps to a distinct peak in mass ratio space. 

\subsection{Comparison to Schneider et al. (2023) estimates}
\label{sec:comparison_to_previous_estimates}

Here, we compare the results of the population synthesis under the \modelMaltsevBalanced model to the predictions from \citet{Schneider_etal.2023_BimodalBlackHole}.
In the mass distribution of individual \acp{BH} (left and middle panels of Fig.~\ref{fig:mass_spectrum_lmhm}), the peaks at $\sim\,9\Msun$ and 19\,\Msun, and the empty gap between $\sim11-16$\,\Msun, are in close agreement with the predictions. 

At the low-mass end, the \acp{BH} with mass $\Mbh<8\,\Msun$ are almost exclusively fallback \acp{BH} in the \modelMaltsevBalanced model, while those with $8\leq\Mbh/\Msun<11$ are formed predominately via direct collapse, which is consistent with the predictions from \citet{Schneider_etal.2023_BimodalBlackHole}.
However, fallback \acp{BH} ultimately do not contribute very much to the merging-\ac{BBH} population (see central panel of Fig.~\ref{fig:mass_spectrum_lmhm}), because of the increased likelihood of binary disruption due to mass loss and natal kicks during the core collapse.
After reweighting by the cosmic \ac{SFH}, we find that \acp{BBH} with at least one fallback \ac{BH} make up just over 2\% of the overall merger rate at current detector sensitivity.

In \citet{Schneider_etal.2023_BimodalBlackHole}, the chirp-mass distribution was predicted to be trimodal as a natural consequence of ``mixing'' pairs of \acp{BH} drawn from a bimodal distribution. 
The locations of the three peaks of the \MchirpS distribution (right panel of Fig.~\ref{fig:mass_spectrum_lmhm}) lie at $\sim$ 9, 12, and 20\,\Msun, in agreement with their expectations. 
However, in that study, the authors argued that the middle LM+HM peak should be suppressed evolutionarily, relative to the other two peaks, due to the narrow window of the parameter space in which a very massive primary could have a secondary with a post-accretion mass in the right range to form a \ac{BH} in the lower compactness peak.
We found that the peak is less suppressed than expected because the systems contributing to this peak predominantly come from an as-yet-unconsidered formation channel for \acp{BBH}.
These binaries, intriguingly, experience no \ac{MT} before the formation of the first-born \ac{BH}, and subsequently undergo \ac{CEE} after the expansion of the secondary.  
The primaries are sufficiently massive that they exceed the Humphreys-Davidson luminosity limit \citep{Humphreys_Davidson.1979_StudiesLuminousStars} shortly after the \ac{MS} and experience \ac{LBV} mass loss.
In COMPAS, eruptive \ac{LBV} mass loss is averaged out into a high wind, following the treatment in \citet{Hurley_etal.2000_ComprehensiveAnalyticFormulae}. In practice, this rapidly strips the H envelope while the star is on the Hertzsprung gap, preventing these massive stars from expanding further. 
The formation of Wolf-Rayet + massive stars from close binaries that avoid interaction due to strong \ac{LBV} winds was studied in \citet{Vanbeveren.1991_EvolutionMassiveClose}, however, to our knowledge this has not been previously explored in the context of \ac{BBH} progenitors. 
This new \ac{LBV} channel is discussed in more depth in Sect.~\ref{sec:impact_lbv_winds}.

Furthermore, the rise of the third peak (HM+HM) in the \modelMaltsevBalanced model starts at $\MchirpS\approx17-18\,\Msun$, which is somewhat higher than the value of $13-14\,\Msun$ expected in \citet{Schneider_etal.2023_BimodalBlackHole}.
For nearly equal mass \acp{BBH}, the chirp mass is less than the component mass for either \ac{BH}, but if the mass of the lower mass \ac{BH} is $\lessapprox75\%$ that of the higher mass \ac{BH}, the chirp mass exceeds the lower mass.
Thus, in pairings of two HM \acp{BH}, the chirp mass is bounded from below by the lowest attainable mass of a HM \ac{BH}, roughly $16\,\Msun$, except in cases where both \acp{BH} have masses around this lower limit.
From the population synthesis, we find that such $\approx16+16\,\Msun$ pairings of HM+HM \acp{BBH} are rare evolutionarily (see Fig.~\ref{fig:m1_vs_m2}).

\section{Chirp-mass distribution of detectable BBHs}
\label{sec:detector_frame_chirp_masses}

In this section, we focus on the properties of the \ac{BBH} population after convolving with the cosmic \ac{SFH} model and applying observational selection effects.
Unless otherwise specified, chirp-mass now refers to the post-convolution, source-frame chirp mass, which can be directly compared to the LVK observations.

\subsection{Redshift dependence of the predicted chirp-mass distribution}

\begin{figure*}[hbt!]
\centering
\includegraphics[width=\hsize]
{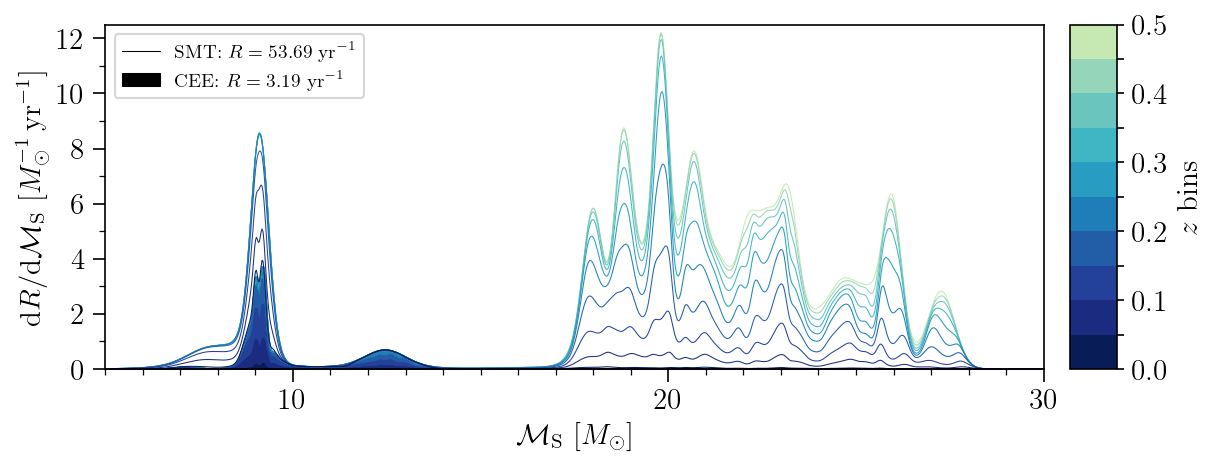}
\caption{
Differential rate of detectable \ac{BBH} mergers $\dv R / \dv \MchirpS$ per unit chirp mass \MchirpS, under the \modelMaltsevBalanced model.
The simulation output was convolved with the default cosmic SFH model described in section Sect.~\ref{sec:calculating_detector_frame_chirp_masses}, assuming approximate O3 sensitivity and accounting for detector bias.
Solid bands show the distribution coming from binaries that experienced CEE, while thin lines show the distribution for SMT binaries. 
The SMT distributions are stacked on top of those for the CEE binaries. 
Colors show the cumulative contribution from different redshift bins, $z \in [0, 0.5]$, beyond which there is a negligible contribution to the expected detections in this lower mass range. 
The total expected detection rate for BBHs from the stable mass transfer and common envelope channels is shown in the upper left corner.
}
\label{fig:default_chirp_mass_d_redshift_bins}
\end{figure*}

In Fig.~\ref{fig:default_chirp_mass_d_redshift_bins}, we show the differential rate of detectable \ac{BBH} mergers $\mathrm{d}R/\mathrm{d}\MchirpS$, split up by mass-transfer history and redshift $z$ bins, up to $z=0.5$, beyond which few \acp{BBH} with $\MchirpS<30\,\Msun$ are detectable.
Notably, we still see the same three peaks as in the pre-convolution \MchirpS distribution (right panel of Fig.~\ref{fig:mass_spectrum_lmhm}), although the highest peak is now much noisier, and the two lowest peaks have a lower overall contribution.
The latter point is perhaps unsurprising, since the detector sensitivity is a strong function of chirp mass, so low-mass, high-redshift binaries are more likely to elude detection.

The contribution from \ac{SMT} to the lowest two peaks is now more similar to that of \ac{CEE}, almost 40\%, despite there being many fewer \ac{SMT} binaries in the lowest peak in the pre-convolution distribution. 
Taken together with the now dominant HM+HM peak, this indicates that \ac{CEE} binaries, although well represented in the initial simulation output, are suppressed during the cosmic \ac{SFH} convolution. 
This is a consequence of the short delay times of \ac{CEE} systems at low chirp masses (see Fig.~\ref{fig:chirp_mass_vs_P}) coupled with the star formation peak at $z\sim2$, roughly 10\,Gyr ago \citep{Madau_Dickinson.2014_CosmicStarformationHistory}.
Low-mass, \ac{CEE} binaries forming at the peak of star formation merge too quickly (at redshifts that are too high) to be observed with current detectors, although a sufficient number of them form at lower redshifts that the LM+HM peak of detectable systems is dominated by \ac{CEE}.

\subsection{Chirp mass predictions for all models}
\label{sec:chirp_mass_predictions_for_all_models}

\begin{figure*}[!htbp]
\begin{subfigure}{\columnwidth}
    \includegraphics[width=\hsize]{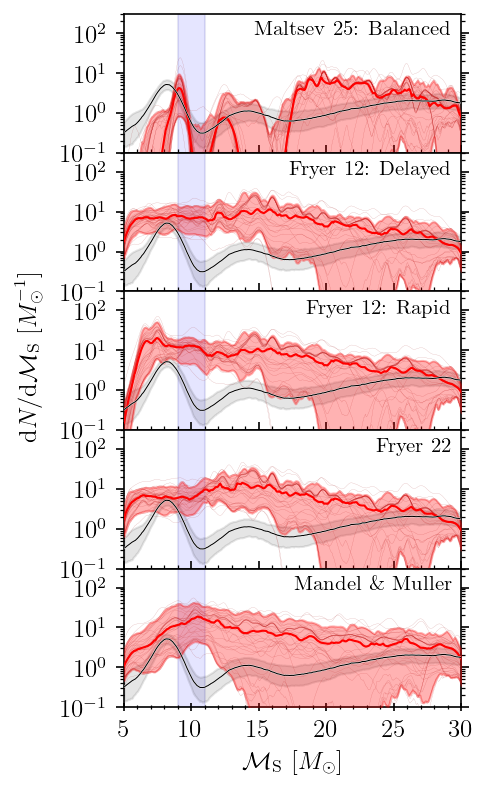}
    \caption{Stellar and binary physics variations}
    \label{fig:number_densities_stbev_vars}
\end{subfigure}
\begin{subfigure}{\columnwidth}
    \includegraphics[width=\hsize]{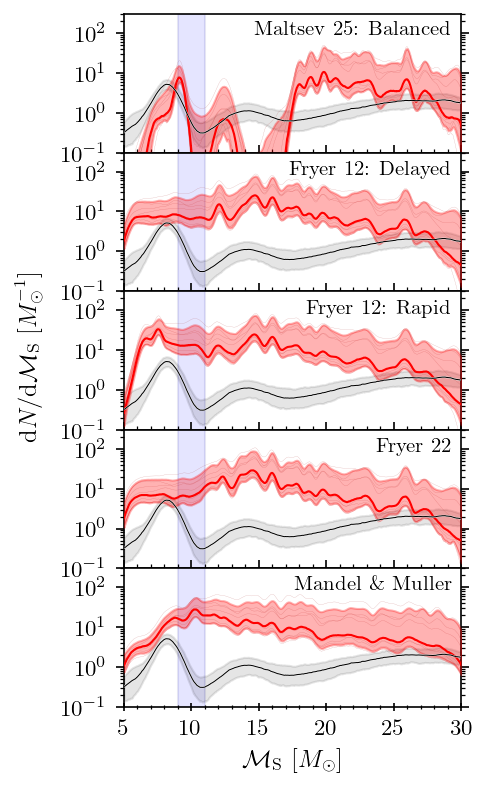}
    \caption{Cosmic SFH variations}
    \label{fig:number_densities_mssfr_vars}
\end{subfigure}
\caption{
Differential number density of detectable \acp{BBH} per unit chirp mass \MchirpS for each supernova model, after a one year effective observing time.
In Fig.~\ref{fig:number_densities_stbev_vars}, we compute the number density for each variation to the stellar and binary physics (see Appendix~\ref{sec:default_and_variations_to_the_stellar_and_binary_physics}), using the default cosmic \ac{SFH}.
The median and 90\% percentile of the range of these density curves are shown as the solid red lines and shaded red bands. 
Individual model densities are shown in light gray.
Fig.~\ref{fig:number_densities_mssfr_vars} shows the same procedure when varying the cosmic \ac{SFH} (see Appendix~\ref{sec:cosmic_star_formation_history}), but fixing the stellar and binary physics to the default model.
The number density for the LVK observations is over-plotted in each row for reference (black lines and gray bands). 
The predicted and observed curves here are constructed using the same adaptive \ac{KDE} setup described in Appendix~\ref{sec:choosing_the_adaptive_kde_bandwidth}, with initial bandwidth $h=0.05\,\Msun$.
}
\label{fig:number_densities_all_models}
\end{figure*}

In Fig.~\ref{fig:number_densities_all_models}, we show the predicted \MchirpS number density \dNdMS for all \ac{SN} models, highlighting the spread in number density predictions due to variations in the stellar and binary physics and cosmic \ac{SFH} physics. 
The specific stellar and binary physics variations are listed in Appendix~\ref{sec:default_and_variations_to_the_stellar_and_binary_physics}, while the cosmic \ac{SFH} variations are described in Appendix~\ref{sec:cosmic_star_formation_history}.
The feature we are most interested in here is how well a given model qualitatively reproduces the observed trimodality in chirp mass, particularly the dip around $10\,\Msun$.
Given the many systematic uncertainties, we do not attempt to quantify the match between the models and the data; our confidence in the models in their entirety does not justify a more formal model comparison at this stage.

In Fig.~\ref{fig:number_densities_stbev_vars}, the number densities are computed for each individual variation to the stellar and binary physics. 
A thorough investigation of each variation and its impact on the \MchirpS number density of \acp{BBH} is left to a future study. 
Instead, we group the number densities for each \ac{SN} model and compute the median and 90\% percentile of the range of the associated curves, which capture the trend and range of predictions allowed for the given assumption on the \ac{SN} physics and \ac{BH} formation. 

It is clear that only the bimodal, \modelMaltsevBalanced model reproduces the observed drop at 10\,\Msun; the alternative (non-bimodal) \ac{SN} prescriptions show a plateau in this region, inconsistent with the LVK data.
However, the \modelMaltsevBalanced model is inconsistent with the data in other regions, such as the underdensities below 8\,\Msun and between 14--17\,\Msun, and the overdensity between 18--25\,\Msun (which is apparent in the other \ac{SN} prescriptions as well). 
Indeed, none of the model variations explored in this study reproduce the observed \MchirpS number density particularly well across the 5--30\,\Msun mass range.  
The same conclusions are apparent when considering the cosmic \ac{SFH} variations in Fig.~\ref{fig:number_densities_mssfr_vars}.
Here, we varied the parameters that describe the cosmic \ac{SFH}, by including the min and max variations described in \citet{vanSon_etal.2023_LocationsFeaturesMass} (see Appendix~\ref{sec:cosmic_star_formation_history}), which capture the range of cosmic star formation histories that still produce a consistent match with the TNG100 simulations used in that study. 
We used the default stellar and binary physics model here (Appendix~\ref{fig:default_chirp_mass_d_redshift_bins}). 

Again, we see that none of the models reproduces the observed \MchirpS number density, and indeed in many cases the cosmic \ac{SFH} variations exacerbate discrepancies in the predicted merger rates, particularly in the alternative \ac{SN} models. 
However, as before, only the \modelMaltsevBalanced model reproduces the sharp drop at 10\,\Msun. 
It is apparent that the cosmic \ac{SFH} variations explored here are unable to explain or induce this sharp drop if it is not present in the underlying \ac{BH} mass landscape.
Comparing the default star formation rate density to those derived from observations \citep[see Fig.~6 in][]{Chruslinska.2024_ChemicalEvolutionUniverse}, the fit from \citet{vanSon_etal.2023_LocationsFeaturesMass} may be somewhat high, however our purpose is not to find an exact fit to the detected chirp-mass distribution, but to explore how variations in these uncertain parameters affect the structural features.

\section{Discussion}
\label{sec:discussion}

\subsection{Support for the bimodal-mass BH model}
\label{sec:support_for_the_bimodal-mass_bh_model}

Now that we have $\approx180$ confident \ac{BBH} detections with a source-frame chirp mass $5<\MchirpS/\Msun<30$, the chirp-mass distribution is starting to show clearer evidence of two larger peaks around $8\,\Msun$ and $27\,\Msun$ and a smaller peak around $14\,\Msun$, with a clear dip at $\sim10\,\Msun$ and a tentative dip around $\sim17\,\Msun$ (see Fig.~\ref{fig:observed_pdfs}). 
The shape of the chirp-mass distribution, and in particular the appearance or absence of the features discussed above, was not reported in \citet{TheLigoScientificCollaboration_etal.2026_GWTC50PopulationProperties}, possibly due to a longstanding convention to focus on the component-mass distributions instead. 
Although their derived component-mass distribution does not show evidence for a bimodal distribution of \ac{BH} masses, this is unsurprising given the poor measurement accuracy for component masses.
While none of the models explored here are fully consistent with the observational data across the entire range of interest (which is to be expected for a population-synthesis pipeline), the positions of the three predicted peaks in the bimodal-mass \ac{BH} model are roughly consistent with the observed peaks, and the agreement in the location of the sharp drop at 10\,\Msun is compelling.

After convolving the population synthesis results using the bimodal \ac{BH} model with the cosmic \ac{SFH}, we see the same three peaks in the \MchirpS distribution (Fig.~\ref{fig:default_chirp_mass_d_redshift_bins}), 
populated by exclusively \ac{SMT} (HM+HM), exclusively \ac{CEE} (LM+HM), or a mixture of both (LM+LM).
This is broadly consistent with previous studies \citep{vanSon_etal.2022_RedshiftEvolutionBinary, Briel_etal.2023_UnderstandingHighmassBinary, vanSon_etal.2023_LocationsFeaturesMass}. 
The interplay of these formation channels was shown to lead to clear structural differences in the number density distributions.

Notably, compared to the location of the observed low-mass peak, the LM+LM peak in the \modelMaltsevBalanced model is consistently translated upward by $\sim1-2\,\Msun$.
The precise locations of the features in these distributions are sensitive to stellar physics assumptions, such as nuclear reaction rates, overshooting, and stellar winds, some of which are difficult or impossible to vary in our setup  
\citep{Schneider_etal.2023_BimodalBlackHole, Laplace_etal.2025_ItsWrittenMassive}.
If these observed \acp{BBH} with the lowest-\MchirpS indeed come primarily from isolated binary evolution with remnant masses that follow a bimodal \ac{BH}-mass model, then the position of this peak offers a rare opportunity to calibrate uncertain stellar physics parameters against gravitational-wave data. 

Overall, the merger rates of the \modelMaltsevBalanced model are about a factor $2-3$ lower than those of the traditional \ac{SN} models.
The integrated number densities from the bimodal models are somewhat more consistent with the $\sim180$ observed events in this mass range (see Fig.~\ref{fig:number_densities_all_models}).
However, we caution against over-interpretation, since the number densities were calculated using a simplistic estimate of LVK sensitivity over a nominal one year of cumulative observing time. 
Since the volume surveyed by the LVK increased by a factor of eight between the O1 and O4 observing runs, a more careful treatment of the absolute rate would require accounting for the changing detector sensitivity and duty cycle of the LVK network.

Lastly, these variations demonstrate that structural features in the chirp-mass distribution are sensitive to the specifics of the adopted binary physics, particularly to the treatment of \ac{SMT}. 
Although we did not attempt a systematic parameter inference study here, these structural features can be used to provide constraints on binary physics and may also help to argue in favor of some \acp{BBH} formation channels over others.

\subsection{Uncertainties in supernova modeling }
\label{sec:uncertainties_supernova_modeling}

Many of the results presented here were based on the assumption that the outcomes of core-collapse \acp{SN} can be tied to the explodability criteria outlined in \citetalias{Maltsev_etal.2025_ExplodabilityCriteriaNeutrinodriven}.
Of course, the specifics of the \ac{SN} modeling, such as the mass and natal kicks of fallback \acp{BH}, are still active areas of research and will inform future iterations of our \ac{SN} model. 
Indeed, the 2D and 3D models of different groups currently do not find agreement on explodability \citep[see, e.g., Fig.~2 in][]{Janka.2025_LongtermMultidimensionalModels}.
It is nontrivial to compare these explodability patterns to the ones obtained from the \citetalias{Maltsev_etal.2025_ExplodabilityCriteriaNeutrinodriven} core-collapse SN recipe, since the latter uses a different set of stellar evolution models \citep{Schneider_etal.2021_PresupernovaEvolutionCompactobject, Schneider_etal.2023_BimodalBlackHole} for mapping out the final fate landscapes as a function of CO-core mass, which have yet not been explored as progenitors of 3D \ac{SN} simulations.

However, our conclusions are somewhat detached from these specifics, and rely only on the following requirements for the explosion modeling:
\begin{itemize}
    \item that this is a threshold process, i.e., stars can only explode if a certain degree of neutrino heating is reached such that the supernova shock can be revived after it initially stalled, 
    \item that this process is determined by the core structure at core collapse, and
    \item that objects in the compactness peak (or a similar peak from a combination of pre-collapse parameters) form direct-collapse \acp{BH}, while objects on either side of the peak form fallback \acp{BH} or \acp{NS}.
\end{itemize}

Any successful neutrino-driven \ac{SN} model (see \citealt{Janka.2017_NeutrinoEmissionSupernovae,Burrows_etal.2024_PhysicalCorrelationsPredictions}) that can satisfy these conditions will reproduce the bimodality in the \ac{BH}-mass distribution.
If preferential \ac{BH} masses exist, as the gravitational-wave data seems to show, it would be natural to expect that these come from the initial conditions for the collapse -- the core structures of their progenitor stars -- rather than from the explosion process itself, which depends on these initial conditions. 
Several recent 3D neutrino-driven \ac{SN} studies have found successful shock revival in high-core mass progenitors, producing fallback \acp{BH} of lower mass than expected if they had directly collapsed. 
Finding consistent explosions of such models would be in contradiction with the large fraction of high-mass \acp{BH} uncovered by the LVK detectors and with possible observational evidence for a lack of high-mass stars producing successful core-collapse supernovae \citep[e.g.,][]{Smartt.2009_ProgenitorsCorecollapseSupernovae,Davies_Beasor.2020_RedSupergiantProblem}.
Therefore, the observed chirp-mass trimodality can be interpreted as a test for recent SN models that find successful explosions of stars with high compactness, such as those which argue that a critically large compactness is not indicative of direct collapse \citep[e.g.,][]{Chan_etal.2018_BlackHoleFormationa,Kuroda_etal.2018_FullGeneralRelativistic,Ott_etal.2018_ProgenitorDependenceCorecollapse,Burrows_etal.2023_BlackHoleFormation,Boccioli_etal.2025_NeutrinoHeating1D,EggenbergerAndersen_etal.2025_BlackHoleSupernovae}.

\subsection{Impact of LBV wind mass loss}
\label{sec:impact_lbv_winds}

Intriguingly, we found that the middle LM+HM peak is primarily formed via a \ac{BBH} formation channel that has not previously received much attention, in which \ac{LBV} winds in the primary deplete the stellar envelope before radial expansion occurs. 
In this \ac{LBV} channel, the first \ac{MT} episode therefore occurs after the formation of the primary, high-mass \ac{BH}. 
While the number of observed events in this middle bump is still too small to provide clear evidence, we speculate that as the number of observed \ac{BBH} mergers grows, the relative height of the middle peak could provide a useful constraint on the importance of the \ac{LBV} channel in forming low-mass ratio LM+HM \acp{BBH}. 

However, it is unclear whether this treatment for \ac{LBV} winds is representative of reality, and thus whether we should trust its implications on the formation of \acp{BBH} which interact only after the first core collapse.
Indeed the \ac{LBV} channel for forming LM+HM binaries relies on the lower-mass secondaries expanding to larger radii than their primaries in order to initiate the first \ac{MT} episode, but the maximum radius of massive stars in COMPAS (based on the stellar tracks from \citealt{Hurley_etal.2000_ComprehensiveAnalyticFormulae}) may be over-estimated by an order of magnitude, artificially enhancing the predicted \ac{LBV} channel \citep{Agrawal_etal.2020_FatesMassiveStars}. 
We do find that when the \ac{LBV} winds are turned off, the suppression of the middle peak is more aligned with the expectations of \citet{Schneider_etal.2023_BimodalBlackHole}, so an observational surplus of \acp{BBH} with chirp mass around 12\,\Msun may indicate that this channel is more relevant than previously appreciated \citep[see also][]{Schneider_etal.2015_EvolutionMassFunctions}. 

\subsection{Comparison to previous studies}
\label{sec:comparison_to_previous_studies}

Our conclusions are broadly consistent with \citet{Disberg_Nelemans.2023_FailedSupernovaeNatural}, who used an analytical approximation to population synthesis tuned to the observed type II \ac{SN} rate, combined with a treatment for \ac{SN} outcomes which is also based on the fits from \citet{Schneider_etal.2021_PresupernovaEvolutionCompactobject}, and compared to an earlier LVK catalog, GWTC-3.
In addition to their simplified population synthesis treatment adapted only for solar metallicity, they used asymmetric Gaussians to approximate the chirp-mass posteriors from the LVK data.
By comparison, we used bootstrapping to sample the LVK chirp-mass posteriors on the most recent GWTC-5.0 dataset, which provides a more accurate representation of the underlying data. 
However, we also considered a more narrow range in chirp masses, between 5 and 30\,\Msun, where we believe the effects of the bimodal \ac{BH} masses are more likely to dominate.
This work could be considered an extension of \citet{Disberg_Nelemans.2023_FailedSupernovaeNatural}, using the flexibility of rapid population synthesis to more carefully determine the range of outcomes that result from fixing different \ac{SN} models and varying the stellar, binary, and cosmic physics.

Recently, \citet{Galaudage_Lamberts.2025_CompactnessPeaksAstrophysical} argued that there is strong evidence for bimodality in the \ac{BH} mass spectrum, albeit not for a full gap-like feature at 10\,\Msun, based on hierarchical Bayesian inference of the component masses of \acp{BBH} observed in the previous LVK catalog, GWTC-3 (but see \citealt{Adamcewicz_etal.2024_NoEvidenceDip}). 
While this study takes a complimentary approach on a larger dataset, we emphasize that a critical difference comes from the choice to investigate either the \ac{BBH} component masses versus the chirp mass.
For a given event, the component masses are inferred from the precisely measured chirp mass together with the component mass ratio, which is typically poorly constrained. 
Even though the uncertainty on the luminosity distance propagates into the source-frame chirp mass, which we use for comparison, the observed trimodal structure occurs in a range of masses where observations are only possible at low redshifts, minimizing the error on the source-frame chirp mass.
As a result, features that are readily present in the chirp-mass distribution may get washed out when considering component masses, thus it is preferable to perform the analysis directly in the space of the observables \citep[see also][]{Tiwari.2024_WhatsBinaryBlack, Tiwari.2025_ExaminingGapChirp,Toubiana_etal.2026_ComparingAstrophysicalModels}.

Very recently, \citet{Galaudage.2026_CompactnessPeaksSubpopulations} extended the analysis of \citet{Galaudage_Lamberts.2025_CompactnessPeaksAstrophysical} to also consider inference on the chirp mass, using the subsequent GWTC-4 catalog. 
They found mild support for a \ac{BH} mass model that has separate low mass and high mass components over a model without these, but did not argue for a full, unpopulated gap.

\subsection{Future directions}
\label{sec:future_directions}

There are a number of important future directions that should be explored to more fully investigate the impact of the bimodal \ac{BH} mass distribution.
The stellar and binary variations explored here are meant to be indicative of the uncertainties in rapid population synthesis, but of course these represent only a small subset of the full range of available parameterizations. 
Moreover, the focus here was on the impacts of the \ac{SN} models specifically, averaging over uncertainties in the other model categories. 
It would be illuminating to dig into these individual variations in greater depth, which we leave to a follow-up study. 

In particular, uncertain wind mass loss prescriptions, and especially \ac{LBV} mass loss, may play a significant role in the progenitor-remnant connection \citep{Sander_Vink.2020_NatureMassiveHelium} (albeit, less so at low-$Z$), and therefore the locations of the \MchirpS peaks. 
Similarly, we did not explore \ac{MT} stability variations here. 
The determination of \ac{MT} stability is another critical uncertainty in binary evolution modeling \citep{Hjellming_Webbink.1987_ThresholdsRapidMass,Soberman_etal.1997_StabilityCriteriaMass,Willcox_etal.2023_ImpactAngularMomentum,Picco_etal.2024_FormingMergingDouble,Iorio_etal.2023_CompactObjectMergers,Tauris_vandenHeuvel.2023_PhysicsBinaryStar,Nelemans_etal.2025_FirstPhaseMass,Klencki_etal.2026_ComfortZonesStars,Briel_etal.2026_CaseCaseDetailed,Temmink_etal.2025_GoodThingsCome}.
Given the competing influence of the \ac{SMT} and \ac{CEE} channels in the shape of the \MchirpS distribution, modifications to the stability boundary will certainly impact the predicted number density of merging \acp{BBH}.

Furthermore, the majority of \acp{BBH} are believed to have formed from low-$Z$ binaries; however, the single and binary-stripped stellar evolutionary models, upon which the \citetalias{Maltsev_etal.2025_ExplodabilityCriteriaNeutrinodriven} supernova prescription is based, were only computed at $\Zsun$ and $\Zsun/10$. 
While our extrapolation variants were able to explore the uncertainty at lower $Z$, a more robust treatment of \ac{BH} bimodality requires more detailed models at these lower metallicities. 
This should not strongly impact our results, in which the low-$Z$ sources are subdominant at the lower redshifts where we are currently sensitive, but may become more relevant with future gravitational-wave detector networks, which are sensitive to higher redshifts and lower-$Z$ sources.
However, regardless of the choice of the extrapolation variant, the bimodality in the \ac{BH} formation landscape is preserved. 
Therefore, unless there is some drastic change in the pattern of occurrence of failed supernova outcomes as a function of progenitor CO-core mass of stripped stars at the lowest metallicities, we can expect the bimodal \ac{BH} mass distribution to remain a robust feature. 

The modeling of the cosmic \ac{SFH} was treated using the parameterized approach described in \citet{vanSon_etal.2023_LocationsFeaturesMass}, based on \citet{Neijssel_etal.2019_EffectMetallicityspecificStar}. 
This approach is straightforward to implement, but as with any parameterized model, may not capture some of the nuance of the true metallicity-specific star formation history. 
Convolving the simulation output with more detailed, non-parameterized models may help to build a more reliable, physically accurate sample of merging \acp{BBH} 
\citep{Mapelli_etal.2017_CosmicMergerRate, Chruslinska_etal.2019_InfluenceDistributionCosmic, Lamberts_etal.2019_PredictingLISAWhite}

Finally, the comparison to the observed data was intentionally approximate, using only a binary mask on the false alarm rate and \pastro for inclusion in our sample, as well as approximating the detector as equivalent to O3 sensitivity for 1\,year. 
Such rough approximations are sufficient for building intuition, and moreover the systematic uncertainties from the binary evolution and cosmic \ac{SFH} dominate over those from the data and detector simplifications.
However, a more robust approach would account for the actual false alarm rate and \pastro values to give more weight to confident binaries in the analysis.
Such enhancements are well understood \citep{Farr_etal.2015_CountingConfusionBayesian} and would allow for a more direct comparison between the predicted \MchirpS number density and the observations.

\section{Conclusion}
\label{sec:conclusion}

In this study, we performed a series of rapid population synthesis simulations of interacting binary stars, varying the physics of binary interactions and the cosmic star formation history, to explore their impact on the properties of merging \acp{BBH}. 
We focused in particular on a treatment of core collapse that is based on recent advances in our understanding of the link between final core structures of stars in isolated binary systems and supernova explosion outcomes \citep{Schneider_etal.2021_PresupernovaEvolutionCompactobject, Schneider_etal.2023_BimodalBlackHole, Laplace_etal.2025_ItsWrittenMassive, Maltsev_etal.2025_ExplodabilityCriteriaNeutrinodriven}.
We studied the effects of this model, which predicts a bimodal mass landscape for newly formed \acp{BH}, on the \ac{BBH} source-frame chirp-mass distribution after accounting for the cosmic \ac{SFH} and selection effects. 
We found that this model qualitatively reproduces features in the observed chirp-mass distribution, particularly, the sharp drop at $\sim10\,\Msun$, while the traditional, non-bimodal \ac{BH} formation models do not.
In this model, the merger rate is also found to be a factor of a few below the traditional rates, in better agreement with the observations. 

The bimodal \ac{BH} mass prescription naturally yields a trimodality in the source-frame chirp-mass \MchirpS distribution, which persists after convolving with the cosmic \ac{SFH}.
The peaks are derived from different pairings of \acp{BH} from either the low-mass (LM) or high-mass (HM) peaks, with the highest peak (HM+HM) forming primarily from systems that experienced \ac{SMT}, the middle peak (LM+HM) forming mostly from systems that experienced \ac{CEE}, and the lowest peak (LM+LM) from a mixture of both. 

The observed trimodality is a robust prediction of the bimodal \ac{BH}-mass model when many other binary evolution parameters are varied. 
Other features, such as the width and height of the individual peaks, appear to be sensitive to specific binary physics parameters, while the locations of the peaks depend on assumptions concerning the internal stellar physics.
However, among the traditional, non-bimodal \ac{BH} mass models, none was able to reproduce the observed trimodal structure, even after accounting for variations in the cosmic \ac{SFH}.
Variations in the assumed cosmic \ac{SFH} model produce overall offsets in the normalization of the chirp-mass distributions, but do not completely erase existing features or create new ones, suggesting that such structure really is a smoking-gun of the underlying stellar and binary physics.
If alternative \ac{BBH} formation channels, such as formation in dense clusters, active galactic nuclei, or multiple stellar systems, do not contribute significantly to the observed rates of mergers in the mass range considered here, then the particular features and shape of the distribution will provide unprecedented constraints on these uncertain stellar and binary evolution parameters and the physics of core-collapse supernovae.

Robust, metallicity-independent features of the \acp{BBH} chirp-mass distribution can be leveraged as standard sirens for cosmological purposes; however, this depends on the capability of the detectors to observe such features out to cosmological distances
\citep{Schutz.1986_DeterminingHubbleConstant, Farr_etal.2019_FuturePercentlevelMeasurement, Farmer_etal.2019_MindGapLocation, Ezquiaga_Holz.2022_SpectralSirensCosmology}.
The features at the lowest masses $\MchirpS\lesssim15\,\Msun$, which are most sensitive to the bimodal \ac{BH} mass prescription, are not currently observed significantly beyond $z\sim0.3$ (see Fig.~\ref{fig:default_chirp_mass_d_redshift_bins}).
Such features may be difficult to use for cosmological inference with current ground-based detectors, even at design sensitivity, although third-generation detectors are expected to observe low-\MchirpS \acp{BBH} out to much greater cosmological distances. 
If indeed the observed features in the LVK data can be robustly connected to the weakly $Z$-dependent bimodality in the \ac{BH} mass distribution, this will provide a redshift-dependent feature that can be used as a standard siren. 

\section*{Data availability}
All processed data, as well as the inference, analysis, and plotting scripts required to reproduce the results and figures in this paper, are available at \url{https://zenodo.org/records/20445516}.

\begin{acknowledgements}

The authors thank the anonymous referee for their thorough feedback, as well as Thomas Janka, Lieke van Son, Floor Broekgaarden, Adam Br\v{c}ek, and Katie Breivik for useful conversations that improved the manuscript. 
RW acknowledges support from the KU Leuven Research Council through grant iBOF/21/084.  PM acknowledges support from the European Research Council (ERC) under the European Union’s Horizon 2020 research and innovation programme (grant agreement No. 101165213/Star-Grasp), and from the Fonds Wetenschappelijk Onderzoek (FWO) senior postdoctoral fellowship
number 12ZY523N.
FRNS acknowledges support by the Klaus Tschira Foundation. This work has received funding from the European Research Council (ERC) under the European Union’s Horizon 2020 research and innovation programme (Grant agreement No.\ 945806) and is supported by the Deutsche Forschungsgemeinschaft (DFG, German Research Foundation) under Germany’s Excellence Strategy EXC 2181/1-390900948 (the Heidelberg STRUCTURES Excellence Cluster).
IM acknowledges support from the Australian Research Council (ARC) Centre of Excellence for Gravitational Wave Discovery (OzGrav), through project number CE230100016. EL acknowledges support through a start-up grant from the Internal Funds KU Leuven (STG/24/073) and through a Veni grant (VI.Veni.232.205) from the Netherlands Organization for Scientific Research (NWO).
Simulations were performed with the \textsc{COMPAS} rapid binary population synthesis code \citep{TeamCOMPAS:Riley_etal.2022_COMPASRapidBinary,TeamCOMPAS:Mandel_etal.2025_RapidStellarBinary} v03.12.04. 
Data analysis and plots were created in python 3.11 \citep{vanRossum.1995_PythonTutorial}
using numpy \citep{Harris_etal.2020_ArrayProgrammingNumPy},
scipy \citep{Virtanen_etal.2020_SciPy10Fundamental}, 
astropy \citep{AstropyCollaboration_etal.2018_AstropyProjectBuilding},
h5py \citep{Collette.2013_PythonHDF5},
and matplotlib \citep{Hunter.2007_Matplotlib2DGraphics}. 

\end{acknowledgements}

\bibliographystyle{aa}
\bibliography{bib}

\begin{appendix}

\section{Choosing the adaptive KDE bandwidth}
\label{sec:choosing_the_adaptive_kde_bandwidth}

\begin{figure*}[!ht]
\centering
\includegraphics[width=\linewidth]{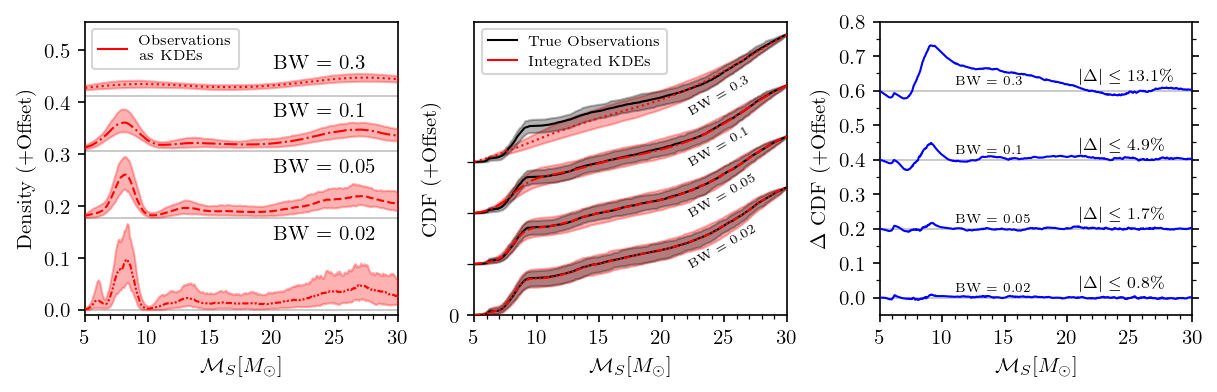}
\caption{
Comparison of different values for the initial \ac{KDE} bandwidth in representing the underlying distribution of observed \ac{BBH} chirp masses. The left plot shows adaptive \acp{KDE} constructed using different initial bandwidth values: central red lines and shaded regions show the median and 90\% percentile of the range of a family of such \acp{KDE} obtained by bootstrapping from the combined LVK event posteriors. The middle plot compares the CDFs of these \acp{KDE} (red) to the true CDF of the observed data (black), with uncertainties estimated through bootstrapping of the observed sample. The right plot shows the difference between the CDFs, together with the maximum absolute difference across this mass range. 
}
\label{fig:bw_comparison}
\end{figure*}

In order to represent the density distributions of the chirp mass populations, we opt for an adaptive \ac{KDE} bandwidth procedure. An adaptive \ac{KDE} adopts variable bandwidths for each kernel, determined by the local density, which is itself calculated from an initial standard \ac{KDE} procedure  \citep{Silverman.1986_DensityEstimationStatistics}. Isolated points, which by definition suffer from low-number statistics, are therefore given broader bandwidths, while points that are tightly clustered have narrower bandwidths and do not ``spill over'' into adjacent low-density regions. As a result, adaptive \acp{KDE} can better reproduce rapidly changing features in an underlying distribution, such as the sharp drop at 10\,\Msun in the observed \ac{BBH} chirp masses. 

For this purpose, we use the python library awkde\footnote{\url{https://github.com/mennthor/awkde}}, which defines the adaptive (unweighted) \ac{KDE} using an initial pilot \ac{KDE},
\begin{equation}
    \tilde{f}(x) = \sum_i \frac{1}{h} \cdot K\left(\frac{x - X_i}{h}\right),
\end{equation}
where $K$ is the kernel (here taken to be a Gaussian) with bandwidth $h$, constructed from samples $X_i$. From this pilot density, a local density factor is defined as 
\begin{equation}
    \lambda_i = \left(\tilde{f}(X_i)/g\right)^{-\alpha},
\end{equation}
where $g$ is the geometric mean of the pilot density,
\begin{equation}
    \log g = \frac{1}{n} \sum_i \log \tilde{f}(X_i),
\end{equation}
and $\alpha$ is a sensitivity parameter satisfying $0\leq\alpha\leq1$. Following \citet[ch.~5.3]{Silverman.1986_DensityEstimationStatistics}, we take $\alpha=0.5$. The adaptive \ac{KDE} then uses the original bandwidth scaled by the local density factor,
\begin{equation}
    \hat{f}(x) = \sum_i \frac{1}{h\lambda_i} \cdot K\left(\frac{x - X_i}{h\lambda_i}\right).
\end{equation}
For our purposes, a weighted adaptive \ac{KDE} would have been desired, to account for the $Z$-reweighting in the simulated samples as well as the probabilities that the observed systems are real and astrophysical. However, as this has not yet been implemented in awkde, we instead used iterative weighted resampling of the target distributions to pre-weight the samples in the adaptive \ac{KDE} (where relevant), and sharp cuts on the astrophysical probabilities for the observed systems, i.e., a step function in \pastro and the false alarm rate.

As with standard \acp{KDE}, this approach is dependent on a subjective choice of the initial bandwidth $h$, which will unfortunately influence the appearance of the final \ac{KDE} and the sharpness of its features. 
Since our analysis relies considerably on visually comparing the presence of such features, we explore several different bandwidth choices for \acp{KDE} constructed from the observed LVK \acp{BBH}. 
In the left plot of Fig.~\ref{fig:bw_comparison}, we construct adaptive \acp{KDE} from the observed LVK \acp{BBH} using 4 different choices for the bandwidth $h$: 0.02, 0.05, 0.1, and 0.3\,\Msun. In practice, a family of adaptive \acp{KDE} is constructed for each bandwidth choice, by bootstrapping samples from the combined posterior of all confident LVK events. 

To benchmark how well each bandwidth choice represents the underlying dataset, we convert the \acp{KDE} distributions to \acp{CDF} by integrating over the domain of the window. 
The \ac{CDF} taken directly from the observations is independent of any \ac{KDE} bandwidth, and so the difference between this and the \ac{KDE}-integrated \acp{CDF} provides a measure of goodness-of-fit. 
The true and integrated \acp{CDF} are shown in the middle plot of Fig.~\ref{fig:bw_comparison}; the 90\% percentile of the range of the observed data is again obtained via bootstrapping. The difference between the medians of these curves is shown in the right plot. 

Of course, the goodness-of-fit can be arbitrarily improved by reducing the bandwidth, as the best representation of the data is a series of delta-functions, so there is some balancing needed to obtain a bandwidth that is sufficiently accurate while remaining useful. 
Critically, the observed \acp{CDF} show a clear plateau at 10\,\Msun, affirming that the apparent, sharp drop in the density distribution is a real feature of the data (and not, e.g., a quirk of the \ac{KDE} bandwidth). 

Here, we demonstrate that the \ac{KDE} with a bandwidth of 0.05\,\Msun has a \ac{CDF} that differs from that of the observations by $\leq1.5\%$ across the mass range of interest. This \ac{KDE} convincingly shows a drop at 10\,\Msun and peaks at 8 and 25\,\Msun, and provides moderate support for a small bump at 13\,\Msun. The broader bandwidths, 0.1\,\Msun and the Silverman Rule of Thumb method, wash out these features and disagree with the \ac{CDF} to a correspondingly higher degree, while the narrower bandwidth of 0.02\,\Msun induces sharper features in the \ac{KDE}, such as the small peak at 6\,\Msun. While we cannot rule out the presence of this smaller peak as distinct from the larger 8\,\Msun peak, it is composed of only two \acp{BBH} systems, and thus dominated by low-number statistics. Choosing such a narrow bandwidth would emphasize these systems as distinct from the neighboring sub-population. Throughout the manuscript, we therefore opt for a bandwidth of 0.05\,\Msun in all 1D chirp-mass \acp{KDE}.

\section{Details of the bimodal prescription}
\label{sec:details_of_bimodal_prescription}

\begin{table}[h]
\caption{Core-collapse outcome boundaries in \Msun, from \citetalias{Maltsev_etal.2025_ExplodabilityCriteriaNeutrinodriven}.}
\centering
\begin{tabular}{ l l l l l }
\hline
& Case A & Case B & Case C & No MT \\
\hline\hline
\MOne   (\Zsun) & 7.4  & 7.7  & 6.6  & 6.6  \\
\MTwo   (\Zsun) & 8.4  & 8.3  & 7.1  & 7.2  \\
\MThree (\Zsun) & 15.4 & 15.2 & 13.2 & 13.0 \\
\hline
\MOne   (\ZsunByTen) & 7.0  & 6.9  & 6.3  & 6.1  \\
\MTwo   (\ZsunByTen) & 7.4  & 7.9  & 7.1  & 6.6  \\
\MThree (\ZsunByTen) & 13.7 & 13.7 & 12.3 & 12.9 \\
\hline\hline
\end{tabular}
\label{tab:maltsev_prescription}
\end{table}

Here, we elaborate further on the interplay between the formalism from \citetalias{Maltsev_etal.2025_ExplodabilityCriteriaNeutrinodriven} for predicting the outcome of core-collapse supernovae and the extrapolation schemes in metallicity that we adopt in our \ac{BH} mass model in COMPAS.
The bimodal \ac{BH} mass prescription supplies values for the failed supernova outcome boundaries \MOne, \MTwo, and \MThree at metallicities \Zsun and \ZsunByTen, for four different evolutionary histories: whether the first interaction of the star as a donor was Case A, Case B, or Case C \ac{MT}, or whether the star experienced no \ac{MT} as a donor (see Sect.~\ref{sec:bimodal_black_hole_mass_model}). 

For a given \ac{MT} history, we thus have boundary values \MOne, \MTwo, and \MThree at the two reference metallicities (see Table \ref{tabl:mssfr_consts}). We then interpolate and extrapolate the metallicity, depending on the chosen extrapolation variant, Optimistic, Pessimistic, or Balanced. From the system metallicity $Z$, we define an effective metallicity,
\begin{equation}
Z_\mathrm{eff} = 
  \begin{cases}
    Z & \mathrm{Optimistic}, \\        
    \min(\max(Z, \ZsunByTen), \Zsun) & \mathrm{Pessimistic}, \\
    \min(\max(Z, \ZsunByFifty), \Zsun) & \mathrm{Balanced}.        
  \end{cases}
\end{equation}
These variants can be accessed in COMPAS by setting the option ``\textsf{--remnant-mass-prescription}'' to \textsf{MALTSEV2024} together with ``\textsf{--maltsev-mode}'' set to one of \textsf{OPTIMISTIC, PESSIMISTIC,} or \textsf{BALANCED} (the latter is the default).

Next, for each $i$ in $\{1, 2, 3\}$, we interpolate in terms of the logarithm of the normalized effective metallicity,
\begin{equation}
    M_i = M_i(\Zsun) + \Big(M_i(\Zsun) -  M_i(\ZsunByTen)\Big) *\log_{10}(Z_\mathrm{eff}/\Zsun),
\end{equation}
which provides the boundary masses \MOne, \MTwo, and \MThree for the system.
The procedure spelled out in Sect.~\ref{sec:bimodal_black_hole_mass_model} determines the outcome for a given CO-core mass from these boundary masses.

\section{Cosmic star formation history}
\label{sec:cosmic_star_formation_history}

\begin{table}[ht]
\caption{Parameters for the cosmic star formation history and their default, max, and min values (see text for explanation), from \citet{vanSon_etal.2023_LocationsFeaturesMass}.  
Values for the SFRD (top) and $dP/dZ$ (bottom) parameters in the cosmic \ac{SFH} calculations. Default values are used throughout, except in Sect.~\ref{sec:chirp_mass_predictions_for_all_models}. 
}
\centering
\begin{tabular}{l l l l}
\hline\hline
\text{Parameter} & \text{Default} & \text{Min} & \text{Max}  \\
\hline
$\aSF$      &  0.02    &  0.01   &  0.03   \\
$\bSF$      &  1.48    &  2.60   &  2.60   \\
$\cSF$      &  4.44    &  3.20   &  3.30   \\
$\dSF$      &  5.90    &  6.20   &  5.90   \\
\hline
$\mu_0$     &  0.025   &  0.007  &  0.035  \\ 
$\mu_z$     &  -0.049  &  0.0    &  -0.5   \\
$\alpha$    &  -1.778  &  -6.0   &  0.0    \\
$\omega_0$  &  1.125   &  0.7    &  2.0    \\
$\omega_z$  &  0.048   &  0.0    &  0.1    \\
\hline
\end{tabular}
\label{tabl:mssfr_consts}
\end{table}

\begin{figure}
\centering
\includegraphics[width=\columnwidth]{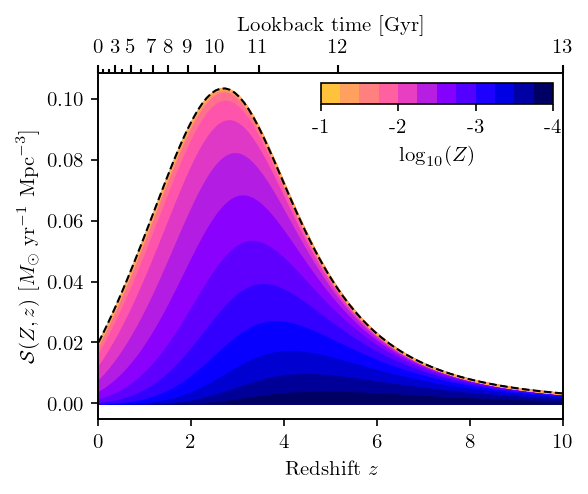}
\caption{
The cosmic star formation history $\mathcal{S}(Z,z)$ as a function of metallicity $Z$ and redshift $z$, using the default values in Table~\ref{tabl:mssfr_consts}. The black dashed line is the total star formation rate density at a given redshift (see Eq.~\ref{eq:sfrd}). 
The lookback time was computed using the cosmological model \textsf{Planck18} in astropy \citep{PlanckCollaboration_etal.2020_Planck2018Results}.
}
\label{fig:SZz}
\end{figure}

To calculate the cosmic \ac{SFH}, we follow the treatment from \citet{vanSon_etal.2023_LocationsFeaturesMass}.
We define 
\begin{equation}
    \mathcal{S}(Z, z) = \mathrm{SFRD}(z) \times \frac{\dv P}{\dv Z}(Z,z).
\end{equation}

The first factor, $\mathrm{SFRD}(z)$, is the total star formation rate density at redshift $z$, per unit time $t$ and per unit comoving volume $V_c$. The SFRD takes the analytical form, 
\begin{equation}
    \mathrm{SFRD}(z) = \frac{d^2 M_\mathrm{SFR}}{dt dV_c}(z) = \aSF \frac{(1+z)^{\bSF}}{1 + [(1+z)/\cSF]^{\dSF}},
\label{eq:sfrd}
\end{equation}
in units of [$\Msun\,\mathrm{yr}^{-1}\,\mathrm{Mpc}^{-3}$] \citep{Madau_Dickinson.2014_CosmicStarformationHistory}.
For the fitting parameters $\{\aSF, \bSF, \cSF, \dSF\}$, we use by default the best fitting parameters from \citet{vanSon_etal.2023_LocationsFeaturesMass} (see Table~\ref{tabl:mssfr_consts}), where the units of \aSF match those of $\mathrm{SFRD}(z)$ and the rest are unitless.

The second factor, $\dv P/\dv Z(Z,z)$ is the probability distribution of the metallicity $Z$ at redshift $z$ that is used in star formation.
Parameterizing the metallicity distribution at a given redshift as a skewed-log-normal, and assuming that the mean metallicity is log-linear with redshift, metallicity distribution with redshift takes the form (see \citet{vanSon_etal.2023_LocationsFeaturesMass} for a derivation),
\begin{equation}
  \frac{\dv P}{\dv Z}(Z, z) = 
    \frac{2}{\omega(z)Z} \phi\left(\frac{\ln Z  - \xi(z)}{\omega(z)}\right) \Phi\left(\alpha\frac{\ln Z - \xi(z)}{\omega(z)}\right),
\end{equation}
where $\phi(\cdot)$ follows the log-normal distribution,
\begin{equation}
    \phi\left(x\right) = \frac{1}{\sqrt{2\pi}}\exp\Big(-\frac{x^2}{2} \Big),
\end{equation}
and $\Phi(\cdot)$ is a factor to account for asymmetry in the metallicity distribution,
\begin{equation}
    \Phi\left(x\right) 
    = \frac{1}{2}\left[ 1+\text{erf}\left(\frac{x}{\sqrt{2}}\right)\right].
\end{equation}

The free parameters then are the skewness $\alpha$, the scale
$\omega$, and the location $\xi$. The skewness parameterizes the deviation from a log-normal distribution, such that $\alpha>0$ skews toward lower metallicities, and is assumed to be constant with $z$.
The scale $\omega$, by contrast, is assumed to evolve log-linearly with redshift,
\begin{equation}
    \omega(z) = \omega_0 \cdot 10^{\omega_z z}.
\end{equation}

The location $\xi$ is related to the mean $\mu(z)$ of the metallicity distribution via
\begin{equation}
    \xi(z) = \ln\Big(\frac{\mu(z)}{2\Phi(\beta\omega)}\Big) - \frac{\omega^2(z)}{2}, 
\end{equation}
with
\begin{equation}
   \beta := \frac{\alpha}{\sqrt{1+\alpha^2}},
\end{equation}
where $\mu(z)$ is also log-linear with redshift,
\begin{equation}
    \mu(z) = \mu_0 \cdot 10^{\mu_z z}.
\end{equation}
Collectively, the total cosmic \ac{SFH} $\mathcal{S}(Z,z)$ is parameterized by 9 variables, which are listed in Table~\ref{tabl:mssfr_consts} together with the best-fit values and uncertainties derived in \citet{vanSon_etal.2023_LocationsFeaturesMass}. 
These are the default values we use when convolving with cosmic \ac{SFH}. 
For convenience, we present the default cosmic \ac{SFH} graphically in Fig.~\ref{fig:SZz}. 

In Sect.~\ref{sec:chirp_mass_predictions_for_all_models}, the cosmic \ac{SFH} parameters are varied according to the method outlined in \citet{vanSon_etal.2023_LocationsFeaturesMass} (see their Table~2).  
The SFRD parameters $\{\aSF, \bSF, \cSF, \dSF\}$ are simultaneously set to either the max or min values while all $dP/dZ$ parameters remain fixed to their default values; or the $dP/dZ$ parameters are set to either the max or min values one parameter at a time while all other parameters remain fixed to default values.  

\section{BBH mass ratio distribution}
\label{sec:bbh_mass_ratios}

\begin{figure}[hbt!]
\centering
\includegraphics[width=\columnwidth]{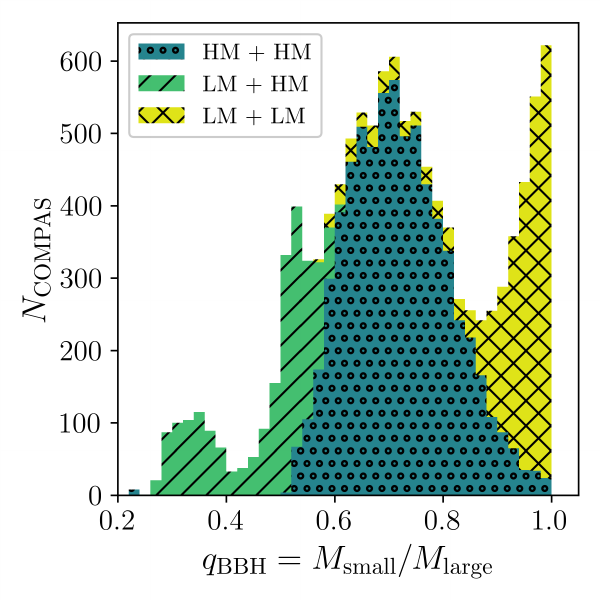}
\caption{Mass ratio \qbbh for \acp{BBH}, defined such that $\qbbh \leq 1$ to be consistent with the \ac{LVK} definition.
These results are from the binary population synthesis output, prior to convolution with the cosmic \ac{SFH} or the application of selection effects.
Colors and hatching correspond to \acp{BBH} composed of \acp{BH} which are both low mass (LM+LM), both high mass (HM+HM), or one of each (LM+HM).
}
\label{fig:qBBH}
\end{figure}

In Fig.~\ref{fig:qBBH}, we show the mass ratio distribution of merging \acp{BBH}, defined as the ratio of the least massive to the most massive \ac{BH}. 
Colors distinguish \acp{BBH} that contain two low mass \acp{BH} (LM+LM), two high mass \acp{BH} (HM+HM), or one of each, LM+HM. 
Each pairing dominates different regions and peaks in mass ratio space, with the lowest mass binaries, LM+LM, peaking at $\qbbh\sim1$ and the highest mass binaries, HM+HM, peaking closer to $\qbbh\sim0.7$.
Comparing against Fig.~\ref{fig:m1_vs_m2}, we identify the lowest mass ratio binaries -- LM+HM -- as those which experienced mass ratio reversal. 

\section{Default and variations to the stellar and binary physics}
\label{sec:default_and_variations_to_the_stellar_and_binary_physics}

Here, we describe the specific variations to the stellar and binary physics that were explored in this study. 
Except where specified, all of these variations (38 in total) were applied together with each of the 5 \ac{SN} models described in Sect.~\ref{sec:bimodal_black_hole_mass_model} and Sect.~\ref{sec:alternative_sn_models}.
For the case of the \citet{Maltsev_etal.2025_ExplodabilityCriteriaNeutrinodriven} prescription, only the default \modelMaltsevBalanced model was used.
Additionally, we included one-off simulations for the \modelMaltsevOptimistic and \modelMaltsevPessimistic models using only the default stellar and binary physics. 
Similarly, we explored four variations to the \ac{SN} fallback fraction \ffb applied only to the \modelMaltsevBalanced model, since the other \ac{SN} models explicitly specify this value. 
These additional variations are included with the other \modelMaltsevBalanced variations in the top row of Fig.~\ref{fig:number_densities_stbev_vars}.

Thus in total, we have run 196 simulations.
While it is impossible to explore every parameter variation and combination in the rapid population synthesis toolbox, those listed here are intended to capture some of the major uncertainties in the field, to highlight the general trend of predictions when only the \ac{SN} model is fixed.
Although we do not comment in depth on the individual model variations, all simulations are available on Zenodo\footnote{https://zenodo.org/records/20445516}, including the equivalent versions of Figs.~\ref{fig:mass_spectrum_lmhm}, \ref{fig:chirp_mass_vs_P}, \ref{fig:m1_vs_m2}, \ref{fig:default_chirp_mass_d_redshift_bins}, and \ref{fig:qBBH}.

\subsection{Fallback \& kicks}

The fallback fraction \ffb (=0.5 by default) represents one of the key uncertainties in the bimodal \ac{BH} mass prescription, so we include alternatives where $\ffb\in\{0,0.25,0.75,1\}$ as well. As stated above, these alternatives are only used together with the \modelMaltsevBalanced model.

The natal kicks that \acp{BH} attain at birth are also a major uncertainty in binary modeling; while observations of isolated pulsars provide strong evidence that many \acp{NS} receive large natal kicks (\citealt{Gunn_Ostriker.1970_NaturePulsarsIii, Lyne_Lorimer.1994_HighBirthVelocities, Hobbs_etal.2005_StatisticalStudy233, Faucher-Giguere_Kaspi.2007_40YearsPulsars,Verbunt_etal.2017_ObservedVelocityDistribution}, though see  \citealt{Tauris_etal.2017_ProgenitorsUltrastrippedSupernovae, Willcox_etal.2021_ConstraintsWeakSupernova,  Valli_etal.2025_EvidencePolarUltralow, Disberg_Mandel.2025_KickVelocityDistribution}), such evidence is much weaker for \ac{BH} kicks and remains an active area of research \citep{Repetto_etal.2017_GalacticDistributionXray,Atri_etal.2019_PotentialKickVelocity,Mahy_etal.2022_IdentifyingQuiescentCompact, Shenar_etal.2022_XrayquietBlackHole, Willcox_etal.2025_BinarityLOwMetallicity,Popov_etal.2025_NatalKicksCompact}. 
By default, we draw natal kicks for \acp{BH} from a single-peak Maxwellian with scale parameter $\sigma=217\,\kms$ \citep{Disberg_Mandel.2025_KickVelocityDistribution} (which is a correction of the commonly used $\sigma_{\mathrm{Hobbs}}=265\,\kms$ based on isolated \acp{NS}, \citealt{Hobbs_etal.2005_StatisticalStudy233}), with \ac{BH} kicks reduced proportionally by the fallback fraction so that direct collapse \acp{BH} receive no kick \citep{Fryer_etal.2012_CompactRemnantMass}.
We include a variation where all \acp{BH} receive no natal kick, regardless of the fallback fraction.

\subsection{Mass-transfer stability}

The uncertain boundary between stable and unstable mass transfer is parameterized with the $\zeta$-prescription from \citet{Hjellming_Webbink.1987_ThresholdsRapidMass}, where $\zeta_{*} = \dv\ln(R) / \dv\ln(M)$ represents the radial response of the star to mass loss at the onset of a mass-transfer event.
This is directly compared to the response of the Roche-lobe to mass loss at the same moment, $\zeta_\mathrm{RL} = \dv\ln(R_\mathrm{RL}) / \dv\ln(M)$, which is an analytical function of the accretion efficiency and angular momentum loss of the \ac{MT} event \citep{Soberman_etal.1997_StabilityCriteriaMass}.
If $\zeta_\mathrm{*} > \zeta_\mathrm{RL}$, the \ac{MT} event is assumed to be stable. 
By default in COMPAS, we use  $\zeta_\mathrm{*}=2$ for \ac{MS} stars,  $\zeta_\mathrm{*}=6.5$ for giant stars with radiative envelopes, and otherwise $\zeta_\mathrm{*}$ follows \citet{Soberman_etal.1997_StabilityCriteriaMass} in treating convective envelope giants as condensed polytropes
\citep{TeamCOMPAS:Riley_etal.2022_COMPASRapidBinary}.
A more detailed treatment of \ac{MT} stability is left to a future study.

\subsection{Common envelope}

If the mass transfer is determined to be unstable, we follow by default the $\alpha_\mathrm{CE}-\lambda$ formalism for \ac{CEE} \citep{vandenHeuvel.1976_LateStagesClose, Webbink.1984_DoubleWhiteDwarfs}. 
The usual efficiency parameter is set to $\alphaCE=1$, but we include variations for $\alphaCE=0.1$ and $\alphaCE=10$ as well. 
Values of $\alphaCE>1$ pertain to energy sources in addition to the orbital energy that contribute to unbinding of the envelope, such as recombination energy \citep{Ivanova_etal.2013_CommonEnvelopeEvolution, Lau_etal.2022_CommonEnvelopesMassive} or feedback from an accreting compact object companion 
\citep{Soker.2004_EnergyAngularMomentum, MorenoMendez_etal.2017_DynamicsJetsCommonenvelope}.
The $\lambda$ parameter follows the implementation from \citet{Dominik_etal.2012_DoubleCompactObjects} based on \citet{Xu_Li.2010_BindingEnergyParameter, Xu_Li.2010_ERRATUMBindingEnergy},
improved as described in \citet{TeamCOMPAS:Mandel_etal.2025_RapidStellarBinary}, which we refer to as $\lambda_\mathrm{Nanjing}$.
We also include a variation to the treatment of \ac{CEE}, \modelTwoStage, based on \citet{Hirai_Mandel.2022_TwostageFormalismCommonenvelope}, which assumes a traditional, rapid common-envelope inspiral only down to the boundary between the convective envelope and radiative inner shell, after which point mass transfer proceeds stably on the longer (thermal) timescale.

\subsection{Stable mass-transfer efficiency}

During stable mass-transfer events, the accretion efficiency $\beta$ is defined as the ratio between the total accreted mass and the total amount of mass lost from the donor via \ac{MT}. 
The \ac{MT} efficiency is generally poorly constrained \citep[see, e.g.,][]{Lechien_etal.2025_BinaryStarsTake}.
A higher $\beta$ means more of the initial stellar mass is retained in the system, resulting in more massive core-collapse progenitors and potentially more massive \acp{BH}. However, $\beta$ also affects the final separation of the binary after the mass-transfer event and therefore the subsequent evolution and likelihood of forming a \tHub-merging \ac{BBH}. 
By default in COMPAS, the accretion efficiency is defined for nondegenerate accretors as
\begin{equation}
    \beta = \betaTherm := 
    \cth \frac{ M_\mathrm{a} / \tau_\mathrm{KH,a} }{ \dot{M}_\mathrm{d}},
\end{equation}
where $\tau_\mathrm{KH,a}$ is the Kelvin-Helmholtz timescale of the accretor, and $\beta$ is bounded between 0 and 1. 
The pre-factor $\cth=10$ by default \citep{Hurley_etal.2002_EvolutionBinaryStars}, and is included to account for the uncertain expansion of the accretor due to rapid accretion, which can lower $\tau_\mathrm{KH,a}$ relative to an equivalent star in thermal equilibrium, and increase the efficiency so long as the expansion does not exceed the Roche-lobe \citep{Lau_etal.2024_ExpansionAccretingMainsequence}. 
We include variations \modelAccEffCOne and \modelAccEffCHundred to model accretors that experience no expansion or significant expansion, respectively. 
We also include variations where $\beta$ is fixed to 1 and 0, corresponding to fully conservative and fully nonconservative accretion. 
If the accretor is spun up to critical rotation during the \ac{MT} event, we assume that angular momentum can couple efficiently between the star and the orbit \citep{Popham_Narayan.1991_DoesAccretionCease}.  
In our default model the accretor will continue to accrete at the same rate, but excess angular momentum from the accreted material is deposited into the orbit \citep{TeamCOMPAS:Mandel_etal.2025_RapidStellarBinary}.
In our treatment, the accretion efficiency for compact-object accretors is always bounded by the Eddington-limited accretion rate (including in the variations with fixed $\beta$), resulting in efficiencies $\beta_\mathrm{CO}\approx0$ except in the case of nuclear-timescale mass transfer.

\subsection{Stable mass-transfer AM loss}

For any stellar material that is lost from the donor but not accreted, the specific angular momentum taken away by this material represents another significant astrophysical uncertainty \citep{Soberman_etal.1997_StabilityCriteriaMass, Pribulla.1998_EfficiencyMassTransfer, MacLeod_Loeb.2020_RunawayCoalescencePrecommonenvelope, 
Willcox_etal.2023_ImpactAngularMomentum, Temmink_etal.2023_CopingLossStability,
Klencki_etal.2026_ComfortZonesStars}.
Similarly to the efficiency $\beta$, variations to the angular momentum treatment affect the orbital evolution during mass transfer, impacting \ac{MT} stability and the final separation.
Common assumptions for the treatment of specific angular momentum include the isotropic re-emission model, in which lost material takes away the specific angular momentum of the accretor, and the outer-Lagrangian mass loss model, in which lost material is assumed to flow out from a nozzle around the Lagrangian point behind the accretor and take away the specific angular momentum at this radius 
\citep{Soberman_etal.1997_StabilityCriteriaMass,Pols.2018_BinaryStarsLecture}.
We approximate the separation between the binary center of mass and this Lagrangian point (which may be L2 or L3 depending on the mass ratio) as $a_\mathrm{L2}/a \approx 2^{1/4}$, which is correct to within 10\% for the relevant mass ratios in this study.

In reality, the specific \ac{AM} loss depends on details of the accretion process which are difficult to model, and likely falls somewhere between these two extremes \citep{MacLeod_Loeb.2020_RunawayCoalescencePrecommonenvelope}.
In this study, we implemented a new prescription for angular momentum loss during stable \ac{MT} in COMPAS.
\ac{AM} loss is parameterized as \fGamma, with \mbox{\fGamma=0} corresponding to isotropic re-emission from the vicinity of the accretor, and \fGamma=1 corresponding to emission from the L2 point, with linear interpolation in $\gamma$, the specific \ac{AM} of the ejected material in units of the specific orbital \ac{AM}, for intermediate values (or extrapolated values, if desired). 

A subtly different parameterization was introduced in \citet{Willcox_etal.2023_ImpactAngularMomentum}, with identical boundary values. However, the interpolation in \citet{Willcox_etal.2023_ImpactAngularMomentum} was linear in the orbital separation $a$ instead of $\gamma \propto a^2$.
Although the differences are minor, we find it desirable to compare our results to the recent work of \citet{Klencki_etal.2026_ComfortZonesStars} who explored intermediate values of the \ac{AM} loss using a linear-in-$\gamma$ approach. 
The new parameterization can be accessed with the COMPAS option \mbox{``\textsf{--mass-transfer-angular-momentum-loss-prescription}''} set to \mbox{KLENCKI\_LINEAR} (cf.~the previous value \textsf{MACLEOD\_LINEAR} for interpolation in $a$). 

In both cases, the value of $\fGamma$ is set via options \mbox{``\textsf{--mass-transfer-jloss-linear-fraction-non-degen}''} or  \mbox{``\textsf{--mass-transfer-jloss-linear-fraction-degen}''} for nondegenerate and degenerate accretors, respectively.
By default, both cases use the isotropic re-emission model, i.e., $\fGammaStar=\fGammaCO=0$, however we include several variations that vary the \ac{AM} loss for both accretor types, with both $\fGammaStar$ and $\fGammaCO$ taking values from {0.0, 0.5, 1.0} (except when $\beta=1$ in which case $\fGammaStar$ is irrelevant and simply set to 0). 
Very high angular momentum loss during \ac{MT} onto a \ac{BH} may be somewhat unrealistic, but we include it here to remain agnostic \citep[though see][]{Lu_etal.2023_RapidBinaryMass}.

\subsection{Main sequence He cores}

We also account for different methods to determine the resulting He-core masses at the end of the \ac{MS}. 
Our default treatment follows \citet{Romero-Shaw_etal.2023_RapidPopulationSynthesis}, where if a star is a \ac{MT} donor on the \ac{MS}, the minimum value of its He-core mass at the end of the \ac{MS} is estimated as the fraction, given by the current fractional age along the \ac{MS}, of the projected final core mass at the end of the \ac{MS} assuming no further mass loss as defined in \citet{Hurley_etal.2000_ComprehensiveAnalyticFormulae}. 
As alternatives, we include the original core mass treatment from \citet{Hurley_etal.2000_ComprehensiveAnalyticFormulae}, model \modelHurleyCores, as well as an updated treatment explored in \citet{Shikauchi_etal.2025_EvolutionConvectiveCore} and \citet{Brcek_etal.2026_EvolutionMassiveMainsequence}, model \modelBrcekCores.
In model \modelHurleyCores, the core mass remains fixed at 0\,\Msun until the end of the \ac{MS} at which point it jumps up discretely to a value determined by the mass at the end of the \ac{MS}. This approach is known to underpredict the final core mass in cases where the star lost significant mass during the \ac{MS}, either via strong winds or binary mass transfer. 
In model \modelBrcekCores, the \ac{MS}-core mass is calculated more carefully, using fits to the convective-core mass from detailed 1D models from \citet{Shikauchi_etal.2025_EvolutionConvectiveCore} with adjustments for mass loss due to winds or mass transfer from \citet{Brcek_etal.2026_EvolutionMassiveMainsequence}.

\subsection{Winds}
Finally, we explore several variations to the treatment of stellar winds in different evolutionary phases.
For hot (e.g., OB) stars, we follow by default the prescription outlined in \citet{Vink_Sander.2021_MetallicitydependentWindParameter}, but include as variations the models from \citet{Bjorklund_etal.2023_NewPredictionsRadiationdriven} and \citet{Krticka_Kubat.2018_GlobalHotstarWind}.
Winds from cool, massive stars (e.g., red supergiants) by default follow \citet{Decin_etal.2024_ALMADetectionCO}, with variations from \citet{Vink_Sabhahit.2023_ExploringRedSupergiant} and \citet{Kee_etal.2021_AnalyticDustindependentMassloss}.
For naked Helium (e.g., Wolf-Rayet) stars, we follow \citet{Sander_etal.2022_OriginImpactWolfrayettype} by default, with a variation from \citet{Shenar_etal.2019_WolfrayetBinariesNitrogen}.

\end{appendix}
\end{document}